\documentclass[twocolumn]{aastex62}
\pdfoutput=1 %for arXiv submission
\usepackage{amsmath,amstext}
\usepackage[T1]{fontenc}
\usepackage{apjfonts} 
\usepackage[figure,figure*]{hypcap}
\usepackage{wrapfig}
\usepackage{makecell}
\usepackage{gensymb}
\usepackage{longtable}
\usepackage{threeparttable}

\shorttitle{Orbital Motion of GSC 6214-210 \MakeLowercase{b}}
\shortauthors{Pearce et al.}

\begin{document}
\newcommand{\s}[1]{\textsubscript{#1}}
\newcommand{\Msun}{ M\(_\odot\)}
\newcommand{\Mjup}{\mbox{M$_{\rm Jup}$}}
\newcommand{\chisquared}{$\chi^{2} \;$}
\newcommand{\plusminus}[1]{$\pm \; #1$}
\newcommand{\rowgroup}[1]{\hspace{-1em}#1} % Indents items in a "rowgroup" in a table
\newcommand{\abs}[1]{\lvert#1\rvert}

\title{Orbital Motion of the Wide Planetary-Mass Companion GSC 6214-210 \MakeLowercase{b}: No Evidence for Dynamical Scattering}

\author{Logan A. Pearce}
\affiliation{Department of Astronomy, University of Texas at Austin, Austin, TX, 78712, USA}
\author{Adam L. Kraus}
\affiliation{Department of Astronomy, University of Texas at Austin, Austin, TX, 78712, USA}
\author{Trent J. Dupuy}
\affiliation{Gemini Observatory, Northern Operations Center, 670 N. A'ohoku Place, Hilo, HI 96720, USA}
\author{Michael J. Ireland}
\affiliation{Research School of Astronomy and Astrophysics, Australian National University, Canberra, ACT 2611, Australia}
\author{Aaron C. Rizzuto}
\affiliation{Department of Astronomy, University of Texas at Austin, Austin, TX, 78712, USA}
\author{Brendan P. Bowler}
\affiliation{Department of Astronomy, University of Texas at Austin, Austin, TX, 78712, USA}
\author{Eloise K. Birchall}
\affiliation{Research School of Astronomy and Astrophysics, Australian National University, Canberra, ACT 2611, Australia}
\author{Alexander L. Wallace}
\affiliation{Research School of Astronomy and Astrophysics, Australian National University, Canberra, ACT 2611, Australia}

\begin{abstract}
Direct-imaging exoplanet surveys have discovered a class of 5-20 \Mjup\space substellar companions at separations >100 AU from their host stars, which present a challenge to planet and star formation models.  Detailed analysis of the orbital architecture of these systems can provide constraints on possible formation mechanisms, including the possibility they were dynamically ejected onto a wide orbit.  We present astrometry for the wide planetary-mass companion GSC~6214-210\,b (240 AU; $\approx$14 \Mjup) obtained using NIRC2 with adaptive optics at the Keck telescope over ten years.  Our measurements achieved astrometric uncertainties of $\approx$1 mas per epoch. We determined a relative motion of $1.12 \pm 0.15$~mas~yr$^{-1}$ (0.61 $\pm$ 0.09 km s$^{-1}$), the first detection of orbital motion for this companion.  We compute the minimum periastron for the companion due to our measured velocity vector, and derive constraints on orbital parameters through our modified implementation of the Orbits for the Impatient rejection sampling algorithm.  We find that close periastron orbits, which could indicate the companion was dynamically scattered, are present in our posterior but have low likelihoods.
For all orbits in our posterior, we assess the detectability of close-in companions that could have scattered GSC~6214-210\,b from a closer orbit, and find that most potential scatterers would have been detected in previous imaging.
We conclude that formation at small orbital separation and subsequent dynamical scattering through interaction with another potential close-in object is an unlikely formation pathway for this companion.  We also update stellar and substellar properties for the system due to the new parallax from \textit{Gaia} DR2.

\end{abstract}


%%%%%%%%%%%%%%%%%%%%%%%%%%%%%%%%% introduction %%%%%%%%%%%%%%%%%%%%%%%%%%%%%%%%%%%%%

\section{Introduction} \label{sec:intro}

The growing population of diverse exoplanetary systems provides opportunities to test theories of star and planet formation, particularly as planets are discovered at the extrema of parameter space.
The population of directly imaged sub-stellar companions near the deuterium burning limit ($\lesssim$13 \Mjup) at very wide separations ($>$100 AU) from their host stars, referred to as wide planetary-mass companions (PMCs), is well positioned for studying star and planet formation, because their orbits make them relatively easy observing targets and ideal for testing formation routes.  
 
Wide-orbit companions are not common, yet do seem to be a standard outcome of star and planet formation.  Occurrence rates of these objects have been found to be 1--4\% \citep{bowler2016,GSC6214}. 
The dominant formation mechanism for these objects remains unclear.  Fragmentation from the same molecular cloud as the primary is capable of producing appropriately-sized cores, but such cores may struggle to prevent further accretion and instead grow to stellar masses \citep{Bate2012}.  Gravitational instability in protostellar disks is capable of forming fragments of several Jupiter masses at wide separations, and so is a promising pathway for PMC formation, but also requires a mechanism to limit growth to only $\sim$10 \Mjup\; \citep{Kratter2010,Dodson-Robinson2009}.  Fragmentation of a Class II disk, after much of the envelope has been exhausted, is also a viable path, however \citet{Andrews2013} showed this is likely to produce wide planetary-mass companions only in the most massive disks.  Classical planet assembly through core accretion is unrealistic in-situ at hundreds of AU on timescales of the ages of the systems for which PMCs are observed ($\tau$ = 1-10 Myr) \citep{Goldreich2004, Levison2001}.  Gas giant planet formation at wide separations is hampered by exceedingly long timescales (longer than the ages of observed systems) \citep{Dodson-Robinson2009}, although pebble accretion can reduce the timescales to below protoplanetary disk lifetimes out to 100 AU \citep{Lambrechts2012,Johansen2017}, which is feasible for systems like HR 8799 \citep{Marois2010}.  However many wide orbit companions in the planetary mass regime lie well beyond 100 AU.

Finally, classical core accretion could form extremely massive giant planets at close radii, near ice lines of 1-2 AU, which could then be ejected out to wide orbits through dynamical interactions at any point in their lifetimes.  If this model dominated, extremely eccentric orbits would be common among these systems, and other, close in companions of equal or greater mass would be needed in the systems to scatter the PMC to its current radius (e.g. \citealt{Ford2008,Veras}).  Thus the system would need to have formed several extremely massive planets.  

It remains unclear if there is one dominant formation channel for the population as a whole.  Observations are hampered by population statistics: relatively few of these objects are known, and fewer have been observed long enough to detect orbital motion and constrain their orbital parameters.
Full orbit studies are becoming feasible for close-in directly imaged giant planets (e.g., HR8799 bcde: \citealt{Konopacky2016}; $\beta$ Pic b: \citealt{Neilsen2014}; Fomalhaut b: \citealt{Pearce2015}, \citealt{Beust2016}; 51 Eri b: \citealt{Derosa2015}).  Most wide planetary-mass companions, however, are still limited to linear orbit arcs.  

Results from orbit studies of wide companions to date have reached mixed conclusions about formation mechanism because of the wide range of orbital parameters among systems, and sometimes even different conclusions for the same system (e.g., GQ Lup; \citealt{Ginksi2014,Schwarz2016}). The wide range in eccentricities found thus far does not suggest a clear common trend. There appear to be low eccentricities for ROXs~12\,b and ROX~42B\,b \citep{Bryan}, a high eccentricity for PZ Tel \citep{Ginksi2014,Biller2010PZTel}, and a range for the systems analyzed by \citealt{blunt}. Additional studies of PMC system orbital architectures can shed light on the range of possible formation pathways for this population group.

In this paper we test the ejection model for the formation of the wide-orbit substellar companion to GSC~6214-210. We perform a detailed study of the relative astrometry and orbital solutions from imaging data over a nine year observational period.  In Section \ref{sec:properties} we describe previous studies of this system and update system parameters in light of the new parallax measurement from {\sl Gaia} data release 2.  In Section \ref{sec:obs} we describe our observations, and in Section \ref{sec:astrometry}, we outline our astrometric measurement and our detection of orbital motion.  In Section \ref{sec:orbit fitting}, we present our orbit fitting procedure and show the constraints our orbital motion measurement places on the orbit configuration.  We also determine that the existence of another, close-in companion cannot be completely ruled out, but is unlikely.  Finally, in Section \ref{sec:discussion}, we show that the ejection model cannot be ruled out, but in-situ formation is more likely to explain this PMC.

\section{System Properties} \label{sec:properties}
\subsection{History}
GSC 6214-210 is a pre-main sequence star identified as a member of the Upper Scorpius subgroup of the Scorpius-Centaurus OB association in \citet{spectraltyperef}.  They observed the star to have lithium absorption (EW = 0.38 \AA) and H$\alpha$ emission (EW = --1.51 \AA) in their survey of X-ray selected PMS candidates.  
\citet{Bowler2014} determined a spectral type to K5 \plusminus1, and mass of M$_{\star} = 0.9 \pm 0.1$ \Msun\space based on their optical spectrum fit to models adjusted for extinction. The median age for USco of 10 Myr \citep{UScoage,Feiden2016} has been adopted for recent studies of this system.  There is no evidence of the primary being a binary system \citep{K08}.

GSC 6214-210 was discovered to host a wide planetary-mass companion by \cite{GSC6214} in their adaptive optics imaging survey for wide companions in Upper Sco.  They found the companion to be very red ($J-K=1.3$ mag, $K-L=1.05$ mag).  \citet{Bowler2014} measured a spectral type of M9.5 \plusminus1 and model-derived mass of M$_{c} = 14 \pm 2$\; \Mjup for the companion.

\cite{bowler2011} reported evidence of Pa$\beta$ emission from the companion that is lacking from the star, implying the presence of a circumplanetary disk that is actively accreting mass, which may have contributed to the red color observed by \cite{GSC6214}.  \cite{bowler2011} studied the survivability of a circumplanetary disk in planet-planet scattering interactions, and infer from the continued presence of the disk that GSC 6214-210\,b\footnote[1]{There is some disagreement in the literature on the appropriate label for the companion, ``B'' or ``b''.  The Simbad entry for the companion gives the label ``b'', so we adopted the lowercase b for the companion in this work to maintain consistency with Simbad.} 
likely did not form at close radius and scatter to its current location through interaction with another companion.  
ALMA observations of the system in \citet{GSCALMA} found no observable emission above background, which shows the disks around the primary and companion to be low mass ($\lesssim$ 0.05 \Mjup\space for the companion). They concluded that both GSC 6214-210 and its companion are likely at the end stages of formation.

No radial velocity surveys have been conducted to search for other unresolved companions, and past studies did not report any indication of an additional companion.

\subsection{Updated properties}\label{sec:updated properties}
Previous studies of the GSC 6214-210 system adopted the mean distance and age of Upper Sco ($d =$ 145 \plusminus14 pc; $\tau = $ 5-10 Myr; \citealt{UScoage}) for this system.  However, \textit{Gaia} DR2 reported a parallax measurement of $\pi$ = 9.19 \plusminus 0.04 mas \citep{Gaia2018}, which gives a distance ($d =$ 108.8 \plusminus0.5 pc) that is considerably closer than was previously assumed.  The astrometric excess noise on the parallax is 0.0, indicating a good parallax solution.  We therefore update the system properties here.  
 
\textit{Gaia} DR2 reported a proper motion of $\mu$ = (--19.38 \plusminus0.10, --31.25 \plusminus0.07) mas yr\textsuperscript{-1} for GSC 6214-210 \citep{Gaia2018}.  This proper motion and distance places GSC 6214-210 near the edge of proper motion space for the USco subgroup \citep{Rizzuto2011,Wright2018}.  We used the matrices of \citet{Johnson1987} and the \textit{Gaia} proper motion and radial velocity measurements to compute new UVW velocities for the system, obtaining $U$ = --4.53 \plusminus0.83 km s\textsuperscript{-1}, $ V$ = --18.01 \plusminus0.11 km s\textsuperscript{-1}, and $W$ = --4.86 \plusminus0.29 km s\textsuperscript{-1}.  These velocities are still consistent with USco membership, given the internal velocity dispersion of the association from \citet{Rizzuto2011} and \citet{Wright2018}.

\begin{figure}[tb!]
\centering
\includegraphics[width=0.5\textwidth]{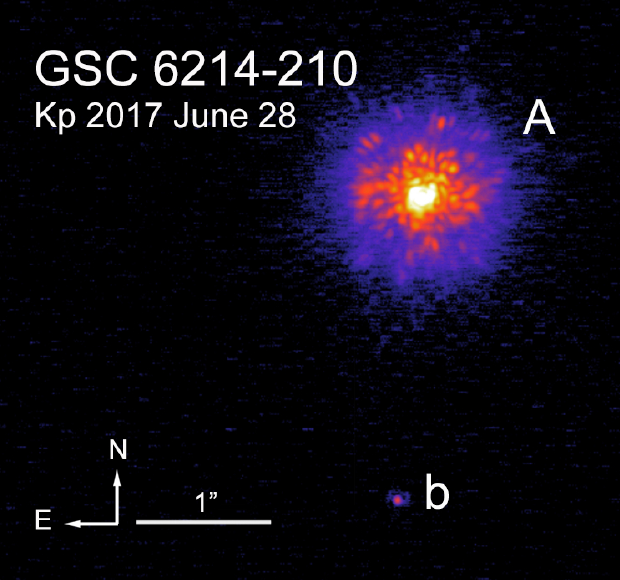}
\caption{\small{Example image of GSC 6214-210 and companion from the 2017 observation epoch.  The companion, labeled ``b'', can be seen at 2.2\arcsec to the south.}} 
\label{fig:GSC6214}
\end{figure}

The revised distance has implications for the primary star age and mass as well.  To update the stellar parameters, we began with the effective temperature and luminosity of T$_{\rm eff \star}$\;=\;4200 \plusminus150 K and L$_{\star}$ = 0.38 \plusminus0.07 L\(_\odot\)\space  reported in \citet{bowler2011}, and calculated the new luminosity to be L$_{\star}$ = 0.221 $\pm$ 0.002 L\(_\odot\) using the same spectral energy distribution calculation from that work.
%then estimated an revised luminosity of $L_{\star}$ = 0.21 \plusminus0.07 L\(_\odot\)\space to be consistent with the revised distance.  
This new luminosity, and the temperature of \citet{bowler2011}, gives a mass and age of M$_{\star}$\;= \;0.80\;\plusminus0.11\;\Msun, $\tau$ = ${{{24}}}\;_{{-{5}}}^{{+{7}}}$ Myr using the evolutionary models of \citet{BHAC15}.  We also compared the temperature and luminosity to the models of \citet{Feiden2016} for young star convection inhibited by magnetic fields, which yielded M$_{\star}$ = 0.78 \plusminus0.06 \Msun, $\tau$ = $41 \pm 10$ Myr.  These ages are in conflict with each other, and highly discrepant with other age determinations of USco members \citep{UScoage,Rizzuto2016}.  However, given the existence of numerous other signatures of youth seen in the primary star (such as lithium absorption, H$\alpha$ emission, and X-ray emission; \citealt{spectraltyperef}), its membership in USco seems to remain secure.
 
For this regime of luminosity and temperature, model-predicted masses depend strongly on age. Adopting a prior informed by the age distribution of the broader population therefore can clarify the likely system parameters and the degree of tension between observations and expectations. The ages suggested for the subgroups of Sco-Cen members span 5-20 Myr \citep{UScoage,Rizzuto2016,PecautMamajek2016}, so we also estimate the age and mass after imposing linear-uniform priors over $5 < \tau < 20$ Myr and (to broadly allow any possible mass) $0.2 <$ M $< 1.4$ \Msun\space. To implement this prior, we first drew random (M,$\tau$) pairs from the linear-uniform distributions, and then computed a corresponding (L,T$_{\rm eff}$) pair for each (M,$\tau$) pair by interpolating the \citet{BHAC15} and \citet{Feiden2016} models. For each pair, we then computed the likelihood that we would have obtained our inferred values of luminosity (L$_{\star}$ = 0.221 \plusminus0.002 L\(_\odot\)) and temperature (T$_{\rm eff \star}$ = 4200 \plusminus150 K), given our estimated uncertainties, if that pair represented the true mass and age of GSC 6214-210. The likelihood was adopted as a weight for that (M,$\tau$) pair. Finally, we computed weighted means of the age and mass across all pairs of potential masses and radii, yielding M = 0.78 \plusminus0.03 \Msun, $\tau$ = $14 \pm 3$ Myr for the models of \citet{BHAC15} and M = 0.83 \plusminus0.02 \Msun\space and $\tau$ = 16.9 $^{+1.9}_{-2.9}$ Myr for the models of \citet{Feiden2016}.  The magnetized models have been shown to more accurately predict mass and age for USco \citep{Feiden2016,Rizzuto2016}, so we adopted this mass and age for the primary.

\begin{deluxetable}{cccc}[tb!]
\tablecaption{{System Properties for GSC 6214-210}\label{table:updates}}
\tablehead{\colhead{Property} & \colhead{Previous Value} & \colhead{Updated Value} & \colhead{Ref} }
\startdata
\multicolumn{4}{c}{Properties of the Primary} \\
\hline
Distance (pc) & 145 \plusminus14 & 108.8 \plusminus0.5 & 1,2\\
Proper Motion  & $\mu_{\alpha}$=-19.6 \plusminus1.1, & $\mu_{\alpha}$=-19.38 \plusminus0.10,  & 3,2\\
(mas yr\textsuperscript{-1}) &$\mu_{\delta}$=-30.4 \plusminus1.1 & $\mu_{\delta}$=-31.25 \plusminus0.07 & \\
Luminosity (L\(_\odot\)) & 0.38 \plusminus0.07 & 0.221 \plusminus0.002 & 4\\
Mass (\Msun) & 0.9 \plusminus0.1 & 0.83 \plusminus0.02 & 5\\
Age (Myr) & 10 & 16.9 $^{+1.9}_{-2.9}$ & 6,7\\
Teff (K) & 4200 \plusminus150 & -- & 4\\
SpT & K5 \plusminus1 & -- & 5\\
\hline
\multicolumn{4}{c}{Properties of the Companion} \\
\hline
Luminosity  & -3.1 \plusminus0.1 & -3.35 $\pm$ 0.02 & 4\\
(log(L/L\(_\odot\))) & & & \\
Age (Myr) & 10 & 16.9 $^{+1.9}_{-2.9}$ & 6,7\\
Mass (\Mjup) & 14 \plusminus2 & 14.5 \plusminus2.0 & 5\\
%Teff (K) & 2200 \plusminus100 & 2050 \plusminus100 & 4\\
%SpT & M9.5 \plusminus1 & -- & 5\\
\enddata
\tablereferences{\small{(1) \citet{deZeeuw1999}; (2) \citet{Gaia2018}; (3) \citet{Zacharias2017UCAC5}; (4) \citet{bowler2011}; (5) \citet{Bowler2014}; (6) \citet{UpperSco}; (7) \citet{Feiden2016}}}
\end{deluxetable}

\citet{bowler2011} used the substellar companion's observed spectral energy distribution to estimate the bolometric luminosity of the companion, log(L/L\(_\odot\)) --3.1 \plusminus0.11 dex.  Given the updated distance measurement, we recalculate this value to be $log(L/L_\odot) =$ --3.35 \plusminus0.02 dex. Adopting the age of $\tau$ = 16.9 $^{+1.9}_{-2.9}$ Myr for the GSC 6214-210 system, we have compared the age and luminosity to the brown dwarf evolutionary models of \citet{Baraffe2003evolutionarymodels} that incorporate updated BT-Settl atmospheres \citep{Allard2012_BTSettl}, as well as to the models of \citet{Burrows1997}. We find model-predicted masses of M$_{c}$ = 14.0 \plusminus0.6 \Mjup\space and M$_{c}$ = 14.9 \plusminus0.2 \Mjup\space respectively.  
To reflect both estimates while still conveying the broad uncertainties in substellar evolutionary models, we adopt a mass of M$_c$ = 14.5 \plusminus2.0 \Mjup\space for the companion for this study.  

Updated system properties are listed in Table \ref{table:updates}.

%%%%%%%%%%%%%%%%%%%%%%%%%%%%%%%% observations %%%%%%%%%%%%%%%%%%%%%%%%%%%%%%%%%%%%

\section{Observations}\label{sec:obs}

We used the near-infrared imaging camera NIRC2 coupled with the adaptive optics system \citep{Wizinowich2000} on the Keck-II telescope to obtain high resolution imaging of the GSC 6214-210 system.  We used a total of 77 images in this study, summarized in Table \ref{table:summary}.  Observations from 2008, 2009, and 2010 were previously reported in \citealt{GSC6214}, and were reanalyzed for this study.  All images used in this study were obtained with the K$^{\prime}$ filter and the narrow camera, with adaptive optics in natural guide star mode.  The details of each image are listed in Appendix \ref{images appendix}.  

Each science and calibration frame was linearized in Python using the methodology of the IDL task \textit{linearize\textunderscore nirc2.pro} \footnote[2]{http://www.astro.sunysb.edu/metchev/ao.html} \citep{Metchev2009}.  Science frames were dark-subtracted and flat-fielded using the corresponding dark and flat frames from each epoch.  Bad pixel identifications were adopted from \cite{stellar} and replaced with the median of surrounding pixels.  An annotated and reduced image can be seen in Figure \ref{fig:GSC6214}.

%%%%%%%%%%%%%%%%%%%%%%%%%%%%%% data analysis %%%%%%%%%%%%%%%%%%%%%%%%%%%%%%%%%%%%%

\section{Astrometry} \label{sec:astrometry}

%%% astrometry %%%

\subsection{Relative Astrometry} \label{sec:relative astrometry}
\subsubsection{PSF Fitting}
\begin{figure}
\centering
\includegraphics[width=0.47\textwidth]{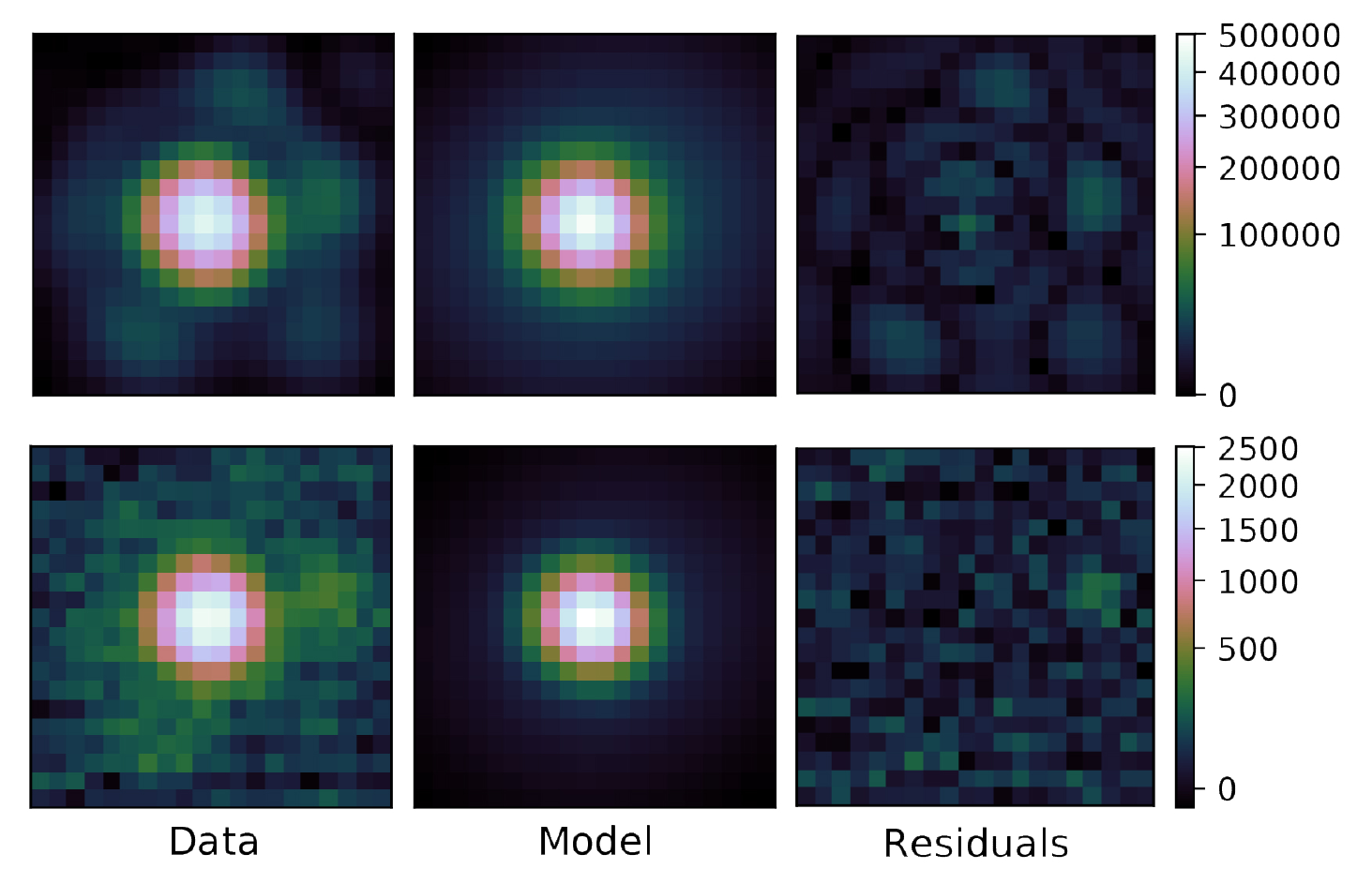}
\caption{\small{Data, model, and residual map of the primary (top) and companion (bottom) for one image from the 2008 dataset.  The model shown is built using the mean values of the parameter chains from the MCMC fit for that image, and is plotted with a square root stretch to emphasize the faint residuals.  The model captures the diffraction-limited core, but did not model the speckle structure.  We used this approach to PSF fitting due to our need to robustly characterize the astrometric error for orbit fitting. We compare our results to results from an empirical model in Section \ref{sec:astrometry results}}}
\label{fig:residuals}
\end{figure}

We determined the relative position of the companion with respect to the host star in each image through a custom Gibbs Sampler Markov Chain Monte Carlo (MCMC) PSF-fitting pipeline.  We modeled the PSF of the primary and companion as the sum of two 2-dimensional Gaussian functions --- one to model the diffraction-limited core and another to capture the wings.  We found that 2 Gaussians were adequate to fully model the symmetric structure in the PSF; extended speckle structure around the primary was sufficiently faint as to not affect the fit, and would be poorly modeled by an azimuthally symmetric function. Robust astrometric error bars were essential for orbit fitting, so we decided that in this case, modeling the PSF analytically was preferable to accurately modeling the speckle noise with empirical PSF templates.  Figure \ref{fig:residuals} displays the data, model, and residual map for the host star and the companion for one image from the 2008 dataset, shown with a square root stretch to emphasize the faint speckle structure.

Our PSF fitter used a Metropolis-Hastings jump acceptance criterion with Gibbs sampling of the 16 model parameters: the $x$ and $y$ position of the primary, $x$ and $y$ position of the companion, amplitude of the primary, amplitude of the companion, amplitude ratio between wide and narrow Gaussian, $x$ and $y$ pixel offset for the center of the wide Gaussian relative to the narrow, background level, standard deviation in $x$ and $y$ direction for both wide and narrow Gaussians, and rotation angles of wide and narrow Gaussians.

We carried out our MCMC fit using the Texas Advanced Computing Center (TACC) Lonestar 5 supercomputer in parallel processing, with each of the 48 cores acting as an independent walker.  We ran each image on Lonestar 5 for 12 hours, generating around 50,000 steps per parameter per walker. We removed the first 13,000 steps to allow for burn-in, which yielded 1.8 million samples per parameter, including $x$ and $y$ pixel position for star and companion.  We computed a Gelman Rubin statistic for each parameter chain for each image and all were below 1.1, which we interpret to indicate sufficient convergence.

We inspected the covariances between each of the fit parameters for each image.  Most joint distributions appeared to be approximately Gaussian with symmetric tails, indicating little covariance between the parameters of the fit.  Some variables were slightly correlated, however none of the variables involved in determining the location of the centroid, our parameter of interest, were correlated in such a way as to bias the location.  We therefore conclude that correlations between model parameters did not influence our astrometric result. 
Separation and position angle were uncorrelated in every image. 

\subsubsection{Astrometric Calibration}

The MCMC chains for the primary and companion pixel positions were then fed into the second phase of our pipeline, which converted each pair of $x$ and $y$ pixel positions to on-sky separation and position angle with appropriate corrections.

To convert the ($x,y$) pixel measurements into on-sky relative positions, we first corrected for optical distortion.  Observations before 2015~Apr~13~UT are interpreted to have a plate scale of 9.952 $\pm \,0.002$ mas pixel$^{-1}$ and PA offset of  --0.252 $\pm$ 0.009\degree\, between the on-chip orientation and on-sky orientation \citep{yelda}, and observations after this date are interpreted to have a plate scale of 9.971 $\pm \,0.004$ mas pixel$^{-1}$ and PA offset of --0.262 $\pm$ 0.022\degree\, \citep{service}, due to the realignment of the NIRC2 camera. FITS header values are set at the start of the exposure, so for images taken in vertical angle mode, we corrected the rotator angles to correspond to the midpoint of the exposure.  We finally applied corrections for differential aberration and atmospheric refraction using the weather data provided in the NIRC2 headers, and computed a relative separation and position angle for each of the 2 million ($x,y$) pixel-space measurements along each MCMC chain.  We computed the final measurement for an image as the mean and standard deviation of the separation and position angle chains.  

To capture the uncertainty in a single epoch, we adopted the weighted average for separation and position angle in each image.  To address systematic errors, we first accounted for the residual error in (x,y) of 1 mas found in both the \citet{yelda} and the \citet{service} distortion solutions by adding a $\sqrt{2}$ mas uncertainty in quadrature to the mean of the positional uncertainties for $\rho$ and $\theta$ within each dither position. We then computed the weighted mean and uncertainty between the dither positions in a single epoch, to reflect that the distortion uncertainty averages down with multiple dither positions.  To that epoch-wide uncertainty we then added in quadrature the pixel scale and orientation correction uncertainties which are common to all dither positions at a given epoch (0.002 mas pixel$^{-1}$ and 0.009\degree\ for the \citet{yelda} distortion solution, 0.004 mas pixel$^{-1}$ and 0.02\degree\ for the \citet{service} distortion solution).  

Error bars in one image are consistent with image-to-image scatter within one epoch, and the scatter around the subsequent linear fit across all epochs (described in Section \ref{sec:astrometry results}).  The reduced chi-squared when combining images within one epoch, and between epochs when doing the linear fit, are close to 1.  Because the companion is detected at low signal-to-noise, the MCMC is constrained by the Poisson error in the companion, and dominates over any systematic error due to PSF offset.

\begin{figure*}[htb!]
\centering
\gridline{\fig{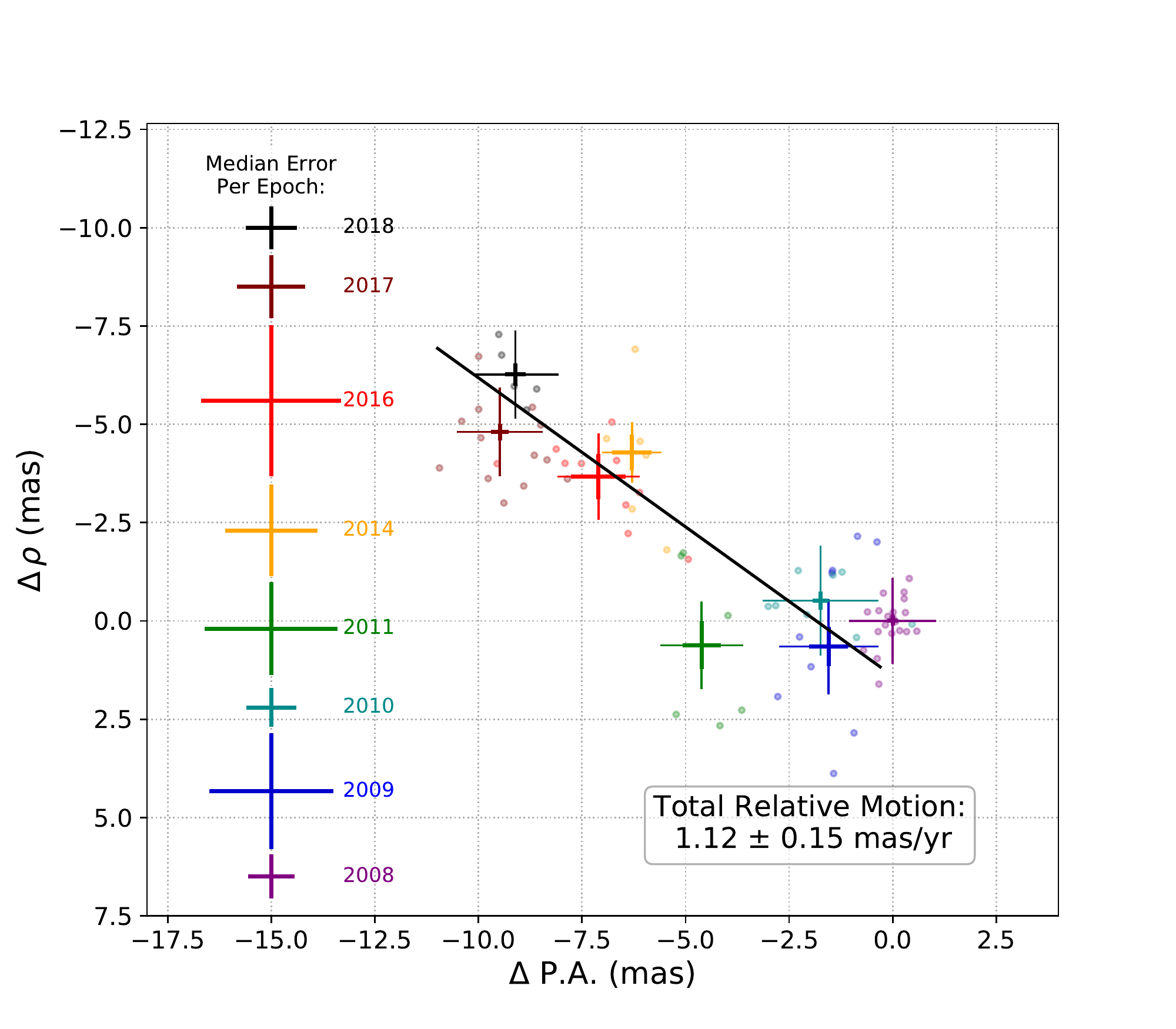}{0.58\textwidth}{(a)}
          \fig{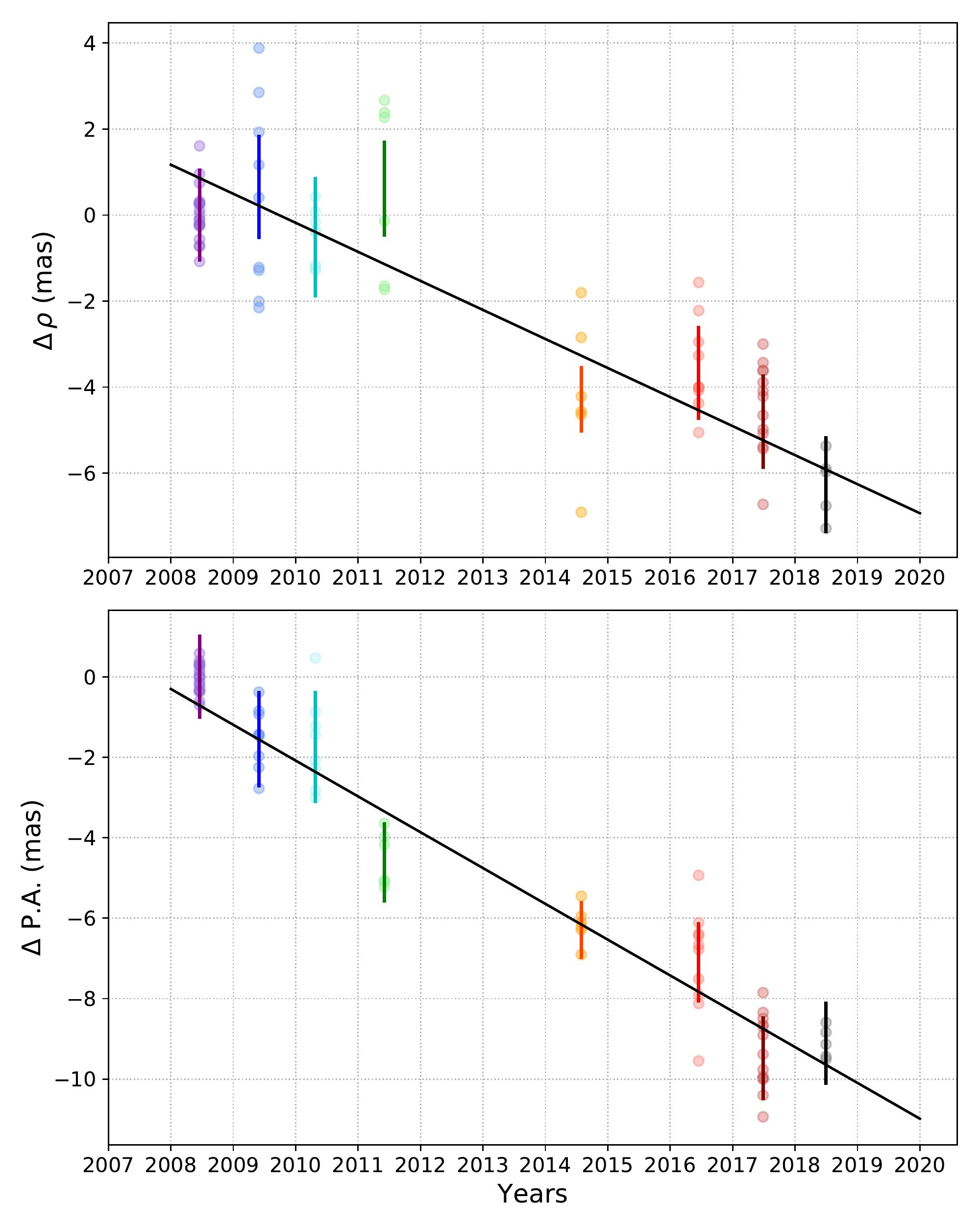}{0.42\textwidth}{(b)}}
\caption{\small{(a): Astrometry results showing a linear trend and orbital motion, plotted as change in angular and radial distance over time. Dots indicate results from each image, and error bars are the mean epoch value.  Thick error bars represent the weighted average uncertainty due to scatter in each epoch, thin error bars show additional systematic uncertainty. Error bars to the left represent median error for a single image. (b): Astrometry and linear fit for each parameter as a function of time.  Error bars represent the total error in each epoch, and match the thin error bars in (a). Deltas reflect change referenced from the average 2008 position, at (0,0).}}
\label{fig:positions}
\end{figure*}

\subsection{Detection of Orbital Motion} \label{sec:astrometry results}

\floattable
\begin{deluxetable}{ccccccccc}
\tablecaption{{Keck/NIRC2 NGS AO Astrometry for GSC 6214-210 b}\label{table:summary}}
\tablehead{
\colhead{Epoch} & \colhead{MJD} & \colhead{Filter} & \colhead{N\s{images}} & \colhead{N\s{Dithers}} & \colhead{t\s{int}} & \colhead{Tracking} & \colhead{Separation} & \colhead{Position Angle}\\
\colhead{} & \colhead{} & \colhead{} & \colhead{} & \colhead{} & \colhead{(sec)} & \colhead{} & \colhead{(mas)} & \colhead{(deg)}
}
\startdata
2008.46 & 54634.34 & K\s{p} & 20 & 2 & 20 & PA & 2201.96 $\pm \,1.09$ & 175.623 $\pm \,0.027$ \\
2009.41 & 54982.38 & K\s{p} & 9 & 4 & 10 & PA & 2202.56 $\pm \,1.20$ & 175.585 $\pm \,0.031$ \\
2010.32 & 55312.61 & K\s{p} & 8 & 4 & 10 & PA & 2201.40 $\pm \,1.40$ & 175.578 $\pm \,0.036$ \\
2011.42 & 55716.34 & K\s{p} & 6 & 3 & 10 & PA & 2202.53 $\pm \,1.12$ & 175.506 $\pm \,0.026$ \\
2014.58 & 56868.27 & K\s{p} & 6 & 1 & 20 & VA & 2197.63 $\pm \,0.64$ & 175.452 $\pm \,0.015$ \\
2016.46 & 57555.47 & K\s{p} & 10 & 1 & 20 & VA & 2198.24 $\pm \,1.10$ & 175.433 $\pm \,0.026$ \\
2017.49 & 57932.31 & K\s{p} & 13 & 1 & 20 & PA & 2197.11 $\pm \,1.13$ & 175.368 $\pm \,0.027$ \\
2018.49 & 58300.37 & K\s{p} & 5 & 1 & 20 & PA & 2195.64 $\pm \,1.12$ & 175.375 $\pm \,0.027$ \\
\enddata
\tablecomments{ 0.026 deg = 1 mas tangential angular distance}
\tablecomments{ N\s{images} is the number of images in each epoch.  N\s{Dithers} is the number of different dither positions in each epoch.  Integration time per image (t\s{int}) is determined as integration time per coadd times the number of coadds. Tracking mode is either Position Angle Mode (PA) or Vertical Angle Mode (VA).}  
\end{deluxetable}

A companion with such a wide separation, and therefore low orbital velocity, requires precise astrometry to distinguish orbital motion from measurement scatter.  We measured linear motion of the companion that is statistically significant above the noise, as reflected in the measurements listed in Table \ref{table:summary} and displayed in Figure \ref{fig:positions}.

We computed a linear fit to the separation and position angle as a function of time, also shown in Figure \ref{fig:positions},  which gives an angular velocity of  $\mu_{\rho} = -0.68 \pm 0.11$ mas yr$^{-1}$ in separation, and $\mu_{\theta} = -0.89 \pm 0.10$ mas yr$^{-1}$ in position angle, with $\mu = 1.12 \pm \,0.15$ mas yr$^{-1}$ in total. The reduced chi-squared statistic of this fit is $\chi^{2} / v$ = 0.89 for 10 degrees of freedom (14 observations fit with 4 parameters), so we conclude that the linear fit is sufficient to describe the data and our errors are of appropriate magnitude. 

We compared our results to astrometry conducted using the empirical PSF-fitting method StarFinder \citep{diolaiti2000}, e.g., as described in \citet{Kepler444},
which found $\mu_{\rho} = -0.54 \pm 0.13$ mas yr$^{-1}$ in separation, and $\mu_{\theta} = -0.98 \pm 0.08$ mas yr$^{-1}$ in PA.  Both are consistent with our result at $\sim$1-$\sigma$, thus uncertainties due to choice of pipeline do not dominate over stochastic uncertainties.

The companion has been well-established to be co-moving with the primary \citep{GSC6214}, and if we treat it as a non-moving background object, 
we would expect to observe a relative motion of 36.77 $\pm$0.13 mas yr$^{-1} $ due to the proper motion of the primary.  This disagrees by $\approx$ 200 $\sigma$ with our measured total velocity.  If the companion were instead another member of USco located at the mean distance of $145 \pm 15$ pc \citep{Rizzuto2011,Wright2018}, it would exhibit a smaller proper motion for the same space velocity, and its relative velocity would also disagree significantly from our measurements.  If we instead treat the companion as completely co-moving, the chi-squared statistic is $\chi^{2}$ = 127.03 ($\chi^{2} / v$ = 12.7), confirming the linear trend is statistically significant above noise.

If the object were in a circular, face-on orbit, the expected velocity at its observed position would be $1.88 \pm 0.05$ km s$^{-1}$ in the plane of the sky. 
At the distance of GSC 6214-210 ($d = 108.8 \pm 0.5$ pc), our measured velocity in the plane of the sky is $0.61 \pm \,0.09$ km s$^{-1}$, indicating that the object is either near apastron, in an inclined orbit, or in a wider orbit which appears close when projected.  
At a projected separation of 240\,au, the only measurements we can make with a 10 year time baseline is projected instantaneous separation and projected orbital velocity. If the separation vector is in the plane of the sky (causing maximum acceleration), then the acceleration would be predicted to be only 6$\times 10^{-4}$\,au\,yr$^{-2}$, resulting in only a 0.05\,mas\,yr$^{-2}$ velocity change. This is well beneath our detection limits, meaning that our data do not constrain acceleration.

Additionally, the observed motion is much smaller than the mean velocity dispersion observed in USco of $\sigma = 1.86 \pm 0.21$ km s$^{-1}$ \citep{Wright2018}. This is further evidence that this is a bound companion and fitting Keplerian orbital parameters to this motion is justified.  With the maximum acceleration of 0.05\,mas\,yr$^{-2}$, deviation from linear motion by more than 3-$\sigma$ (3 mas) would be observable in approximately 11 years.

This linear velocity allows a wide range of possibilities for eccentricity and inclination in our orbit fit but still provides a joint constraint on the posterior distribution of orbital elements.

\subsection{Minimum Periastron}\label{sec:analytical}
If the companion were constrained to a face-on orbit, and thus the most tightly bound, our velocity vector provides the minimum allowed periastron distance, which we computed analytically.  

Using conservation of angular momentum and energy, we can compute that:

\begin{equation}
    L_{obs} = L_{peri}
\end{equation}
\begin{equation}
    E_{obs} = E_{peri}
\end{equation}

where $obs$ denotes values at our observation time, and $peri$ denotes values at periastron.  Thus,

\begin{equation}
    m \,\vec{r_{obs}} \times\ \vec{v_{obs}} = m \, \vec{r_{peri}} \times\ \vec{v_{peri}}
\end{equation}
\begin{equation}
    \frac{-GM_{*}m_{c}}{r_{obs}} + \frac{1}{2}m_{c}{v_{obs}^{2}} = \frac{-GM_{*}m_{c}}{r_{peri}} + \frac{1}{2}m_{c}{v_{peri}^{2}}
\end{equation}

Given that\textbf{ \(\vec{r_{peri}} \times\ \vec{v_{peri}} = \abs{\vec{r_{peri}}}\abs{\vec{v_{peri}}}\sin{\theta}\),} rearranging and substituting for $v_{peri}$, we obtain the quadratic equation

\begin{equation}
    \big(\frac{-GM_{*}}{r_{obs}} + \frac{1}{2}v_{obs}^{2} \big)\, r_{peri}^2\, +\, GM_{*}r_{peri} - \frac{1}{2}\,(r_{obs}\,v_{obs}\,sin\theta)^{2} = 0
\end{equation}

Solving this equation for $r_{peri}$, given the measured mean velocity of $v_{obs} = 0.61$ km s$^{-1}$ and assuming a face-on orbit with $r_{obs}$ = 239.4 AU, we compute a minimum periastron of $r_{peri} =$ 10.0 AU.  Repeating this calculation for a 3-$\sigma$ slower velocity of $v_{obs} = 0.34$ km s$^{-1}$, we compute a periastron of $r_{peri} =$ 3.0 AU.  This gives the minimum allowed periastron to 3-$\sigma$, constrained by our measurement of an orbital velocity vector.

This places an analytical constraint on the minimum, most tightly bound orbit for the companion.  To fully sample the posterior distributions of orbital elements consistent with our astrometric velocity vector, we performed an orbit fitting analysis described below.

%%%%%%%%%%%%%%%%%%%%%%%%%%%% Orbit fitting %%%%%%%%%%%%%%%%%%%%%%%%%%%%%%%%%%

\section{Orbit Fitting} \label{sec:orbit fitting} 

\subsection{Orbits for the Impatient}\label{sec:ofti}

We created a custom orbit fitting algorithm based on the Orbits for the Impatient (OFTI) method of \cite{blunt}.  Briefly, OFTI uses rejection sampling adapted from the methods of \cite{Ghez} to generate probable orbits more quickly than MCMC, but with the same posterior PDF.  Our implementation of OFTI randomly generates four orbital parameters from uniform distributions for eccentricity (\textit{e}), argument of periastron ($\omega$), mean anomaly, and cos(\textit{i}).  The semi-major axis (\textit{a}) for all orbits was initially fixed at 100 AU, and position angle of ascending node ($\Omega$) was initially fixed at 0\degree.  Host star mass and distance were drawn from a Gaussian distribution centered at the updated values from Section \ref{sec:updated properties} value.  Period (\textit{P}) was computed from Kepler's 3rd law for each orbit. Epoch of periastron passage (\textit{T$_{0}$}) was derived from the mean anomaly.  OFTI then scaled \textit{a} and rotated $\Omega$ to match a single observational reference epoch.  We found the reference observation which resulted in the highest acceptance rate, and thus the most efficient, to be one with the minimum error in the astrometry.
%Unlike the \citet{blunt} version of OFTI, which uses previously accepted orbits to determine the optimal epoch to scale to, we selected the middle epoch as the reference observation. 
Astrometric errors were incorporated into the fit by randomly drawing the separation and position angle at the reference epoch from a Gaussian distribution described by our astrometric uncertainties. We placed no additional restrictions on orbital configurations tested by our algorithm.

Like \cite{blunt}, our implementation generated random orbits in batches of 10,000, performed a rejection sampling accept/reject decision on each orbit.
As orbits were generated, the minimum \chisquared found was continuously tracked, and periodically the previously accepted orbits were re-examined with the latest minimum \chisquared to determine if they should have instead been rejected.  This was repeated until a minimum of 100,000 orbits were accepted.

%%%%%%%%%%% orbit results plots and table %%%%%%%%%%%%%%%%%%
\begin{deluxetable*}{cccccccc}[htb!]
\tablecaption{{Summary of Orbital Parameters}\label{table:elements}}
%\tablewidth{0.45\textwidth}
\tablehead{\colhead{Element} & \multicolumn{2}{c}{Circular} & \colhead{Face-on} & \multicolumn{4}{c}{Full Orbit Fit} \\
\cline{2-3}
\cline{5-8}
\colhead{} & \colhead{Inner (blue)} & \colhead{Outer (purple)} &\colhead{} & \colhead{Median} & \colhead{Mode} & 
\colhead{68.3\% Min CI} & \colhead{95.4\% Min CI}  }
\startdata
a (AU) & 246$^{+3}_{-2}$  & 1850$^{+460}_{-380}$ & 128$^{+2}_{-2}$ & 230 & 130 & (125, 300) & (124, 1360)\\
%\hline
e & 0.0 & 0.0 & 0.92$^{+0.02}_{-0.02}$ & 0.66 & 0.87 & (0.51, 0.94) & (0.08, 0.95)\\
%\hline
i (deg) & 106$^{+2}_{-2}$& 96$^{+2}_{-1}$ & 180 & 111 & 105 & (100, 122) & (94, 150)\\
%\hline
%$\omega$ (deg) & \multicolumn{2}{c}{\textbf{Undefined}} & 90$^{+62}_{-61}$ & 83 & 71 & (32,135) & (3,183)\\
$\omega$ (deg) & \textbf{...} & \textbf{...} & 90$^{+62}_{-61}$ & 83 & 71 & (32, 135) & (3, 183)\\
%\hline
$\Omega$ (deg) & 133$^{+1}_{-2}$ & 49$^{+5}_{-5}$ & ... & 53 & 179 & (0, 132) & (2, 180)\\
%\hline
T0 (yr) & \textbf{...} & \textbf{...} & 1000$^{+86}_{-110}$ & 320 & 1300 & (-1000, 1200) & (-20000, 2000)\\
%\hline
Periastron (AU) & ... & ... & 10.3$^{+3.0}_{-2.5}$ & 87.1 & 13.9 & (7.3, 160) &(4.4, 700)\\
\enddata
\tablecomments{Orbital parameter posteriors for a circular orbit fit (e = 0), a face-on orbit fit (i = 180$^{o}$), and a fit with no constrained parameters.  The face-on constraint gave well constrained posterior distributions for all parameters, with a 3-sigma minimum periastron distance of 2.8 AU.  The circular constraint gave two unique orbit solutions, displayed in Figure \ref{fig:constrained fits}, an inner, tighter orbit plotted in blue, and an outer, wider orbit plotted in purple.  Errors on values for the circular and face on constraints represent 68\% confidence intervals.
For the full OFTI parameter fit posterior distributions we report the median, mode, and 68.3\% and 95.4\% minimum credible intervals, with marginal posteriors shown in Figure \ref{fig:GSC6214_countours} and joint distributions displayed in Appendix \ref{corner plot}}
\end{deluxetable*}

\begin{figure}[htb!]
\centering
\gridline{\fig{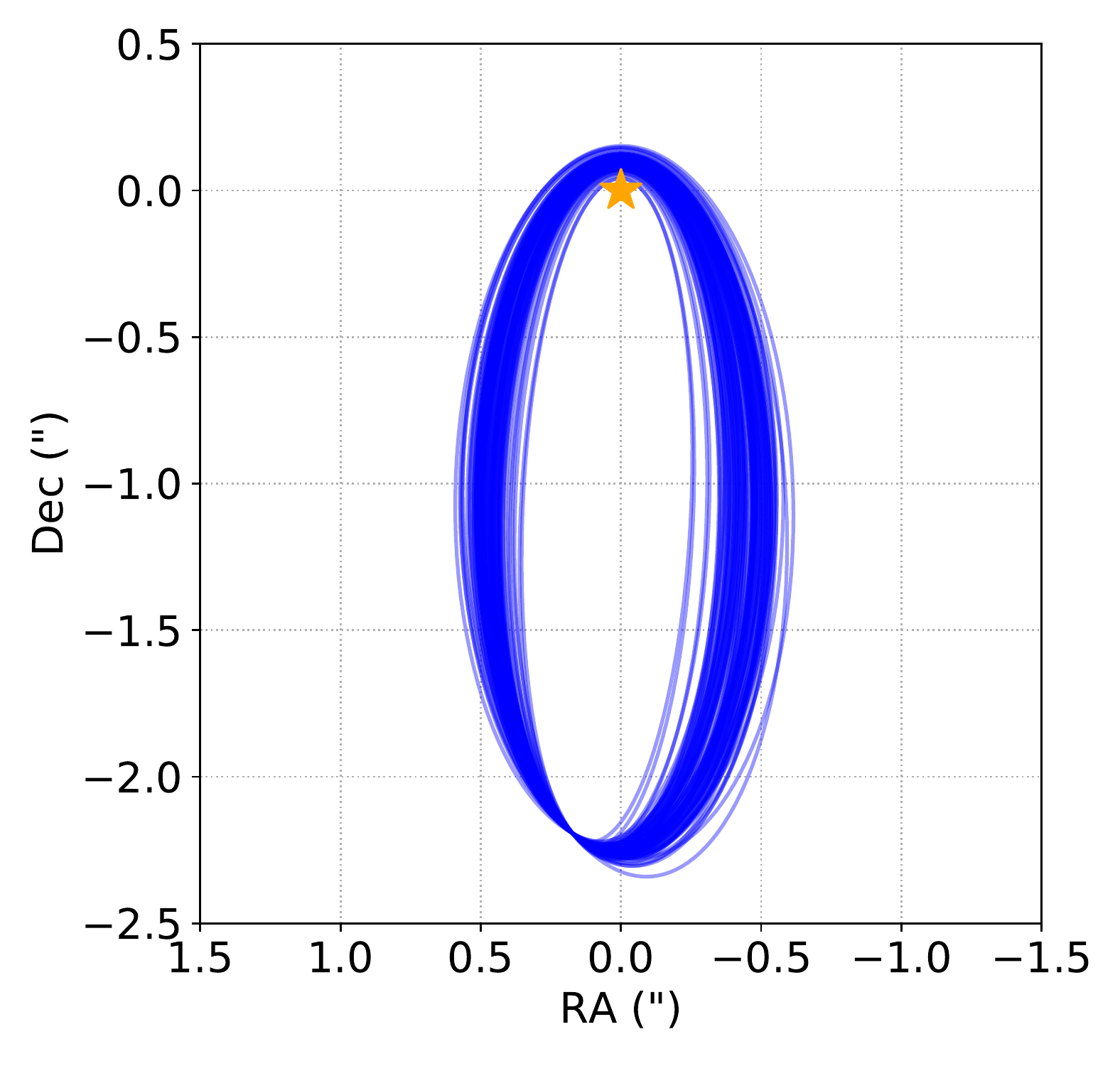}{0.25\textwidth}{(a)}
          \fig{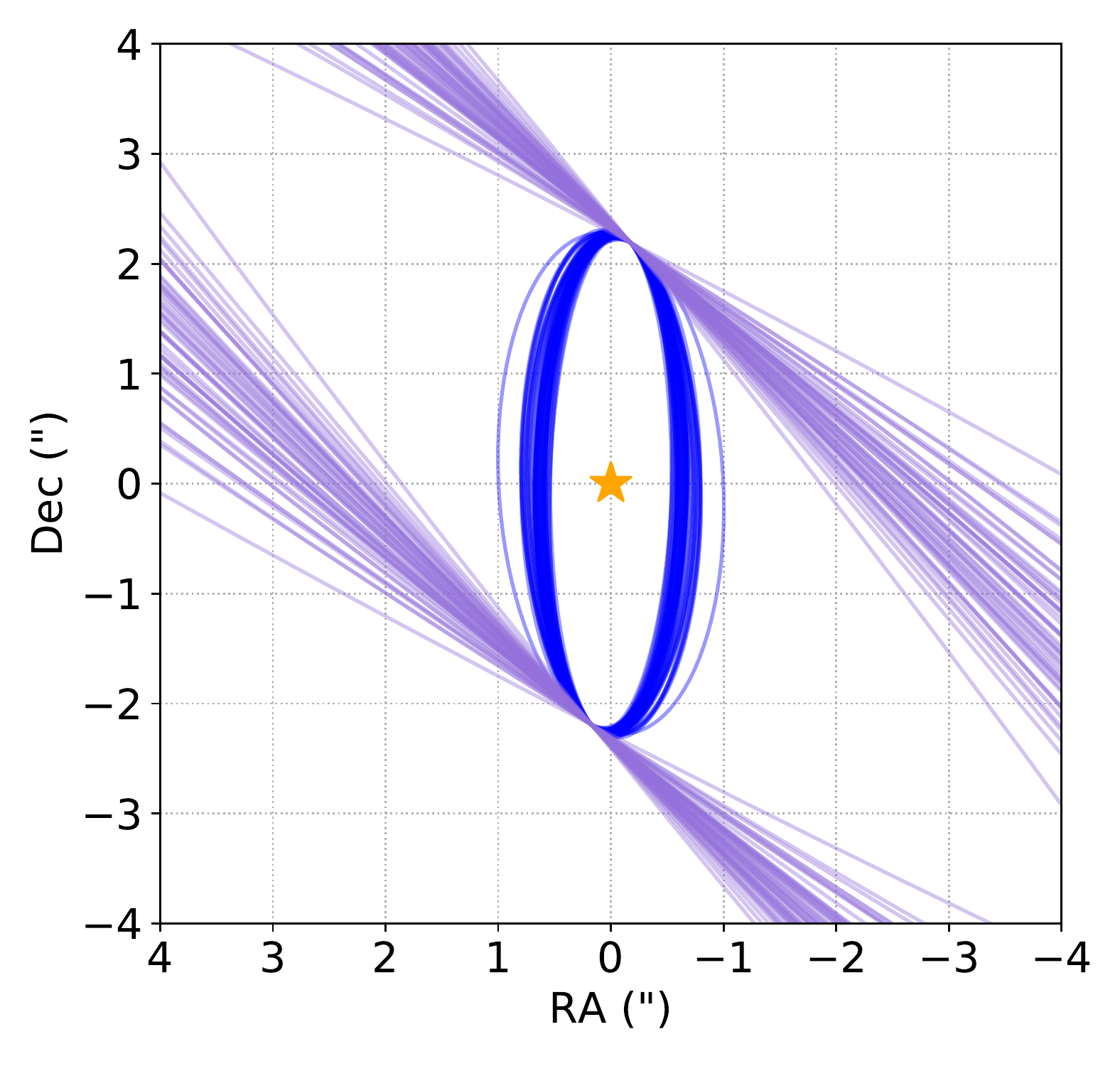}{0.25\textwidth}{(b)}}
\caption{\small{100 orbits randomly selected from the posterior of the two constrained orbit fits shown in Table \ref{table:elements}.  Figure a: Posterior orbits from an OFTI fit with the inclination fixed at a face-on orientation of $i = 180^{o}$.  This is the configuration that places the companion deepest in the potential well.  Figure b: Posteriors from a fit with eccentricity constrained to be circular ($e = 0.0$).  Since the motion in the plane of the sky is well under the circular face-on velocity, two distinct solutions result from this fit: a tighter orbit family plotted in blue, and a wider orbit family plotted in purple.}}
\label{fig:constrained fits}
\end{figure}

%%%%%%% Orbital elements
\subsection{Orbital Elements} \label{sec:orbit results}

\begin{figure*}[htb!]
\centering
\includegraphics[width=1.0\textwidth]{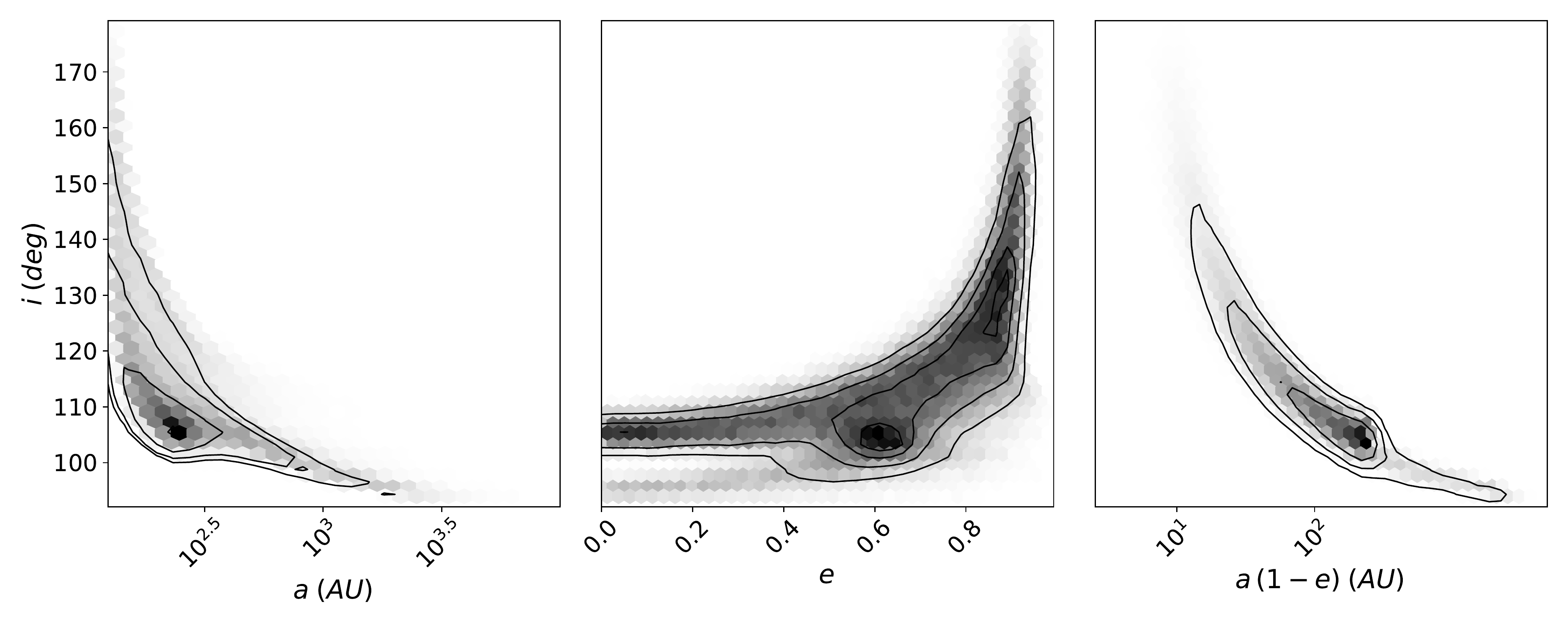}
\includegraphics[width=0.95\textwidth]{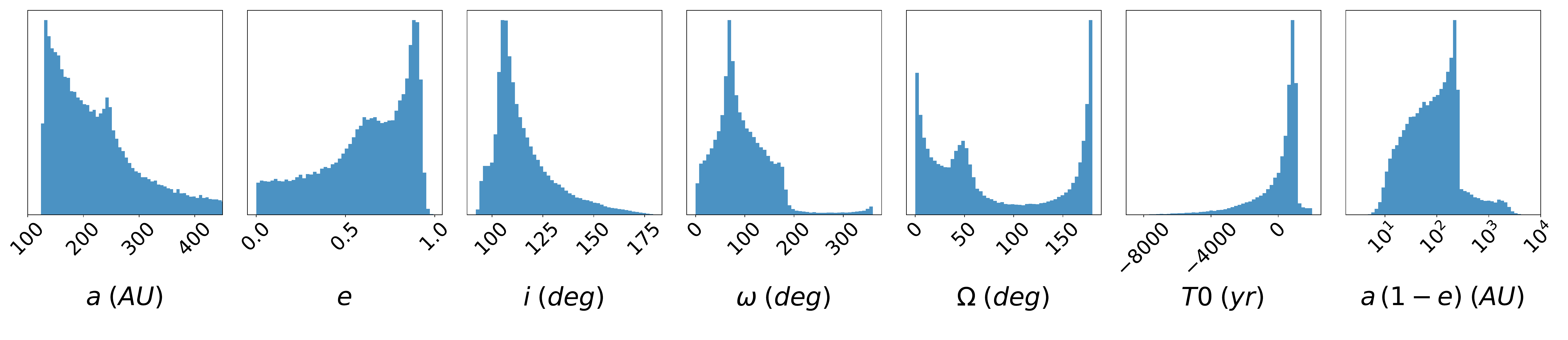}
\caption{\small{OFTI posterior distributions for the orbital elements for GSC 6214-210 b.  Top: Parameter correlations shown as a 2d probability density, with contours indicating 1-, 2-, 3-, and 4-$\sigma$ regions.  These parameters are the most strongly covariant.  Bottom: 1d histograms of the posterior distributions for the fit parameters, plus periastron (defined as $r_{peri} = a \times (1-e) $). In the absence of radial velocity information, there is a degeneracy between solutions with $\Omega$, and consequently $\omega$, which differ by 180$^{o}$.  Here, we follow the convention of selecting the ascending node 0$^{o} \leq \Omega \leq 180^{o}$.  Should future radial velocity measurements indicate $\Omega$ + 180$^{o}$ is the ascending node, then 180$^{o}$ should be added to our measurements of $\Omega$ and $\omega$.  We show corner plot with all of the marginalized two-dimensional credible intervals in Appendix \ref{corner plot}.}}
\label{fig:GSC6214_countours}
\end{figure*}

%%%%%%%%%%%%%%%%%%% Constrained fits %%%%%%%%%%%%%%%%%%
\subsubsection{Constrained fits}
We begin by studying the most extreme limitations our observed velocity vector can place on allowed orbits for the companion.  We performed two OFTI fits with restricted parameters, a face-on fit ($i =$ 180$^{o}$) and a circular orbit ($e = 0.0$).  The results from the two constrained fits are shown in Table \ref{table:elements}.

The face-on fit gave a mean periastron that agrees with our analytic calculation in Section
\ref{sec:analytical}.  When confined to a circular orbit, we find two unique families of solutions, shown in Figure \ref{fig:constrained fits}(b) -- a tighter orbit plotted in blue, and a wider orbit plotted in purple.  Both solutions place constraints on the inclination.  The wide orbit (purple) would place the companion outside the NIRC2 field of view for the majority of the orbit, while the tight orbit (blue) would only hold if we were coincidentally seeing the companion exactly at maximum projected separation. Neither scenario is ruled out by our observations, but they would only hold if we had coincidentally discovered the companion at a very specific epoch in its orbit.

\subsubsection{Full orbit fit}
\begin{figure*}[htb!]
\centering
\gridline{\fig{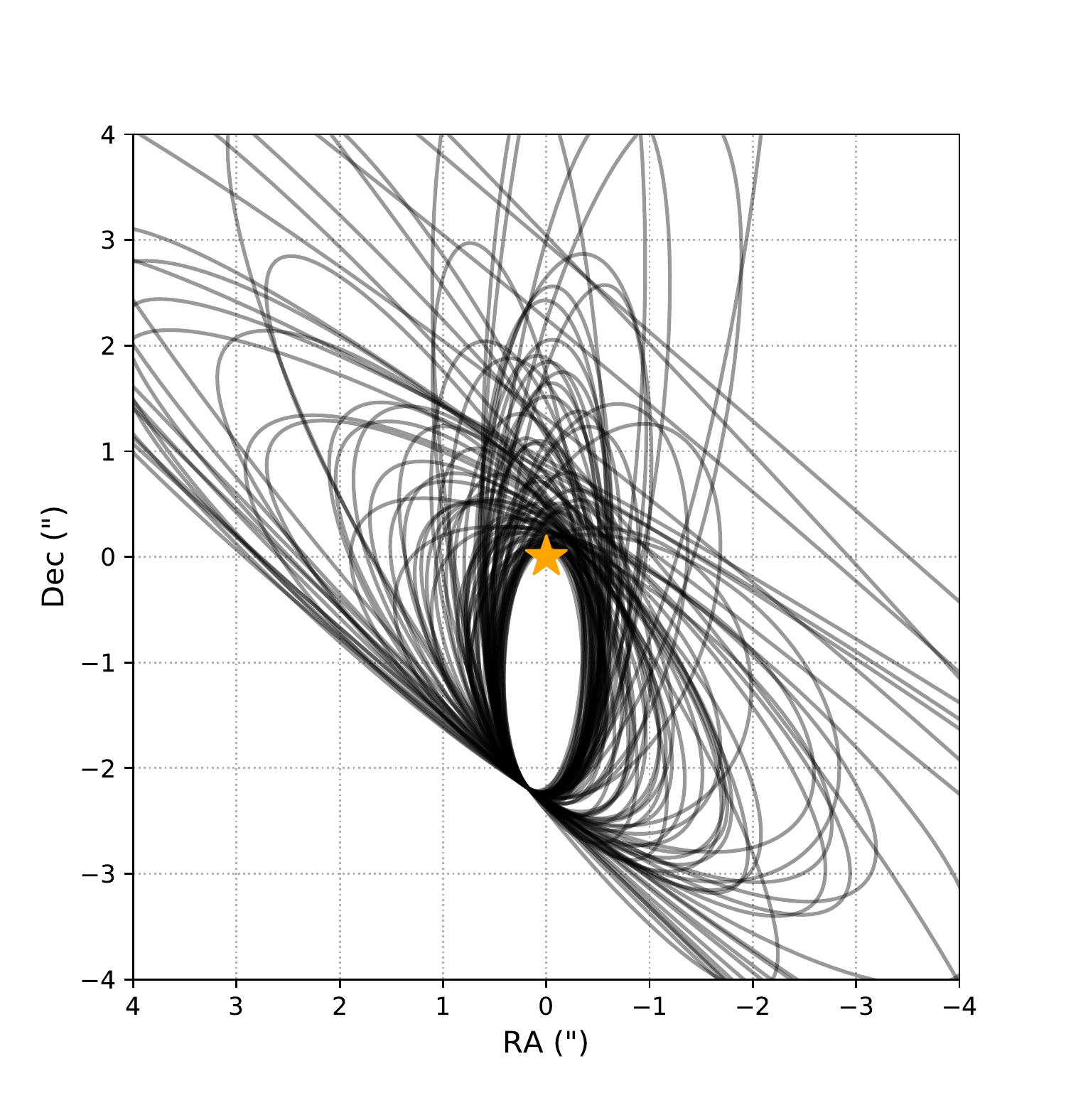}{0.57\textwidth}{(a)}
          \fig{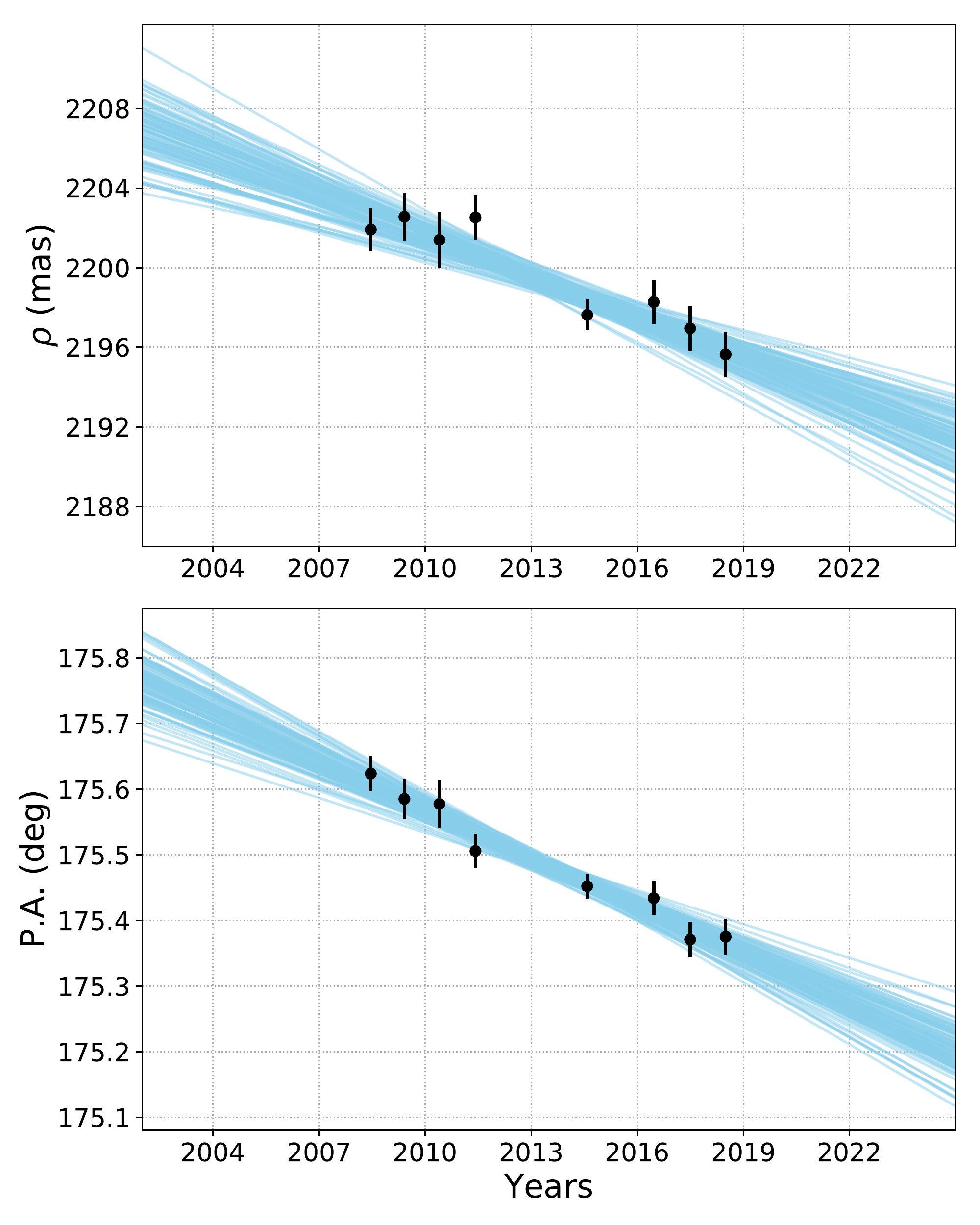}{0.43\textwidth}{(b)}}
\caption{\small{100 orbits randomly selected from the distribution of over 100,000 accepted orbits for GSC 6214-210\,b.  Left: randomly selected orbits around the host star (the yellow star located at 0,0) plotted in right ascension and declination.  Right: the same 100 randomly selected orbits plotted in position angle and separation over time, with the observations from Table \ref{table:summary} over-plotted.  A variety of orbital configurations comprise the posterior of our orbit sample.}}
\label{fig:orbits and pasep}
\end{figure*}

%%%%%%%%%%%%%%%%%%% Unconstrained fit %%%%%%%%%%%%%%%%%%%
We performed an OFTI fit with all free parameters.  Our posterior sample is comprised of 100,000 accepted orbital configurations outputted from OFTI.  
Figure \ref{fig:orbits and pasep} shows a selection of 100 orbits from the sample.  A wide range of orbital parameters were found to fit the data reasonably well, with wide marginal distributions for most parameters.  However, the orbit arc provides a joint constraint on the posteriors of each element.  Figure \ref{fig:GSC6214_countours} displays the joint credible intervals of semi-major axis, eccentricity, and periastron with inclination, and the marginal distributions for all fit parameters and periastron.  Appendix \ref{corner plot} displays the full joint and marginal posterior distributions for the parameters for all orbits accepted by our fitting algorithm.  A large range of values for most parameters were accepted by our OFTI analysis, indicating a wide distribution of possible orbits that satisfy our data.  We report in Table \ref{table:elements} the median, mode, and 68.3\% and 95.4\% minimum credible intervals for the orbital parameters for our full orbit fit.

Finally, because we were interested in the possibility of scattering from close radii, we examined the periastron posterior resulting from each orbit solution in our unconstrained fit posterior.  We find 2\% of orbits having a periastron less than 10 AU.  All orbits with a periastron less than 5 AU have likelihoods less than 5\%, and thus a companion on these orbits would have led to the observed data with a probability less than 5\%.  We therefore conclude that these orbits are not likely to have generated the observations we measured.  

%%% scatterer detection %%%
\subsection{The Detectability of Potential Scatterers}\label{sec:scatterers}

If dynamical scattering were a feasible explanation for the formation of wide PMCs, the observed companion would have been scattered outward by another companion that was likely of equal or larger mass \citep{Veras2004,Veras,Ford2008,Matsumura2017NbodySimulationsofPlanetFormationviaPebbleAccretion}.  No radial velocity surveys to search for unresolved companions have been conducted on this system, and there is no indication of stellar binary companions in spectroscopy on the host star.  The radial velocity scatter and an astrometric jitter term of zero reported by \textit{Gaia} DR2 also argues against the existence of a close-in stellar companion, but is not sufficient to place limits on a substellar companion.  We tested the detectability of another, close-in substellar companion in high-contrast imaging, which could have scattered GSC 6214-210 b to its current wide orbit.

%Our analysis allows us to predict the detectability of potential scattering objects, given existing detection limits for close-in companions.

We modeled the scattering interaction with the simplifying assumption that both objects began on circular orbits at the current periastron distance of each orbit in the posterior sample.  While this is unrealistic physically, it represents the most conservative case because it gives the largest change in orbital energy for the companion, and drives the scatterer down the potential well by the maximum possible amount.  
We determined the change in the companion's orbital energy due to the scattering interaction as
\begin{equation}
\Delta E = -\frac{G M_{\star} m_c}{2} \Big(\frac{1}{a_c} - \frac{1}{r} \Big) 
\end{equation}
where $r$ is the initial radius of the companion's initial circular orbit (equivalent to its current day periastron for any given orbital configuration), and $a_c$ is the companion's final semi-major axis for that orbital configuration.  Assuming all energy lost by the scatterer was transferred to the companion, the final energy for the scatterer is 
\begin{equation}
E_{sc} = -\frac{G M_{\star} m_{sc}}{2r} - \Delta E
\end{equation}
The semi-major axis and eccentricity for the potential scatterer's final orbit are therefore given by:

\begin{equation}
a_{sc} = -\frac{G M_{\star} m_{sc}}{2 E_{sc}}\; ; \; e_{sc} = \frac{r}{a_{sc}}-1
\end{equation}
under the assumption that the original orbital radius ($r$) is the scatterer's new apastron distance, again to conservatively predict the least detectable possible orbital configuration.  
We randomly drew a new orbital inclination for the scatterer and orbital phase for the simulated observation, then applied detection limits to determine if the simulated scatterer would have been detected in imaging data.  We performed this simulation for a scatterer mass equal to the companion (14.5 \Mjup), 1.5 times companion mass (22 \Mjup), twice the companion's mass (29 \Mjup), and five times the companion's mass (72.5 \Mjup).  

As we describe in Appendix \ref{detection limits}, we have determined updated detection limits for additional companions to GSC 6214-210 from each imaging epoch, including those that were previously reported by Ireland et al. (2011), as well as several epochs of non-redundant masking (NRM) interferometry. Given the moderate contrast of GSC 6214-210\,b with respect to its primary star, companions of equal or greater mass (i.e., equal or lesser contrast) would be observable quite close to the primary star, enabling detection in a large fraction of cases. The innermost limit for an equal-brightness companion was found on 2016 Jun 16, where such a companion would have been detected at $\rho >$ 22 AU. Similarly, the limit for a scatterer at 1.5 time the companion's mass was 8.2 AU (on 2011 Jun 04), twice the companion's mass was 1.8 AU (on 2014 Jul 30), while the limit for a scatterer at five times the companion's mass was 1.7 AU AU (on 2014 Jul 30).

To ensure the most conservative estimate, we repeated this analysis for the most energetically limiting case.  The minimum possible initial energy for the scattering object is the one in which results in the final orbit of the scatterer having eccentricity = 1 and semi-major axis = half companion's current periastron distance (and thus apastron at the companion's current periastron), because any angular momentum the scatterer had has been fully transferred to the companion.  Any lower initial scatterer energy could not have produced the system we observe.  This case would result in the scatterer impacting the host star, but it represents the lowest energy limiting case. Thus, regardless of what this initial minimum energy orbit was for the scatterer, we also tested for detectability of a scattering object on this final orbit as the minimum possible configuration for the scatterer.

\begin{figure*}[htb!]
\centering
\includegraphics[width=1.0\textwidth]{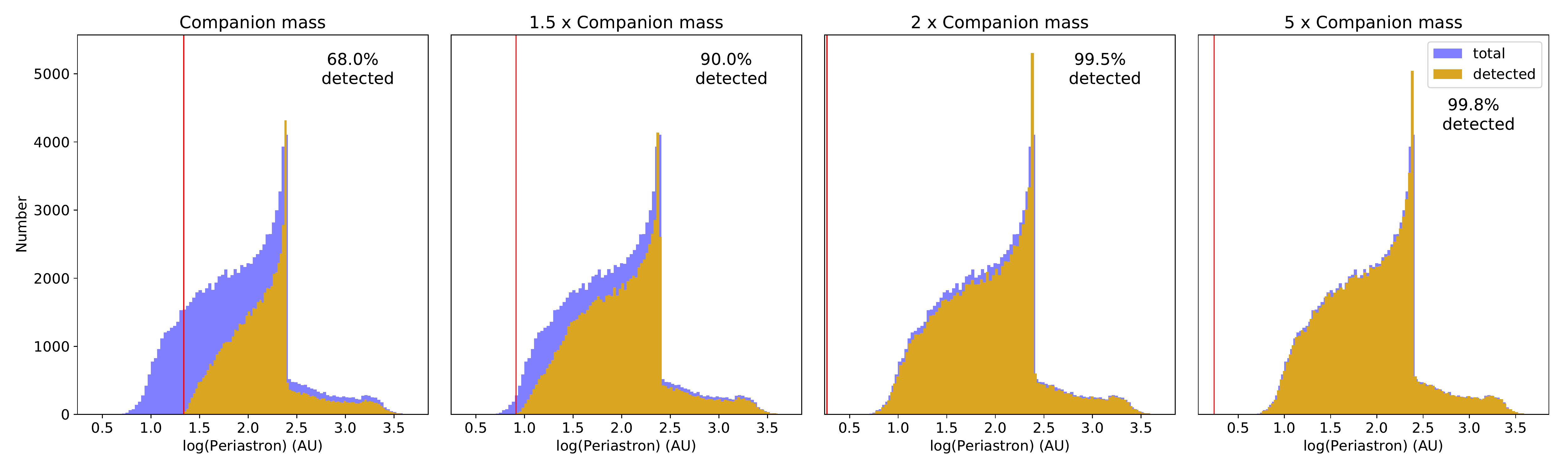}
\includegraphics[width=1.0\textwidth]{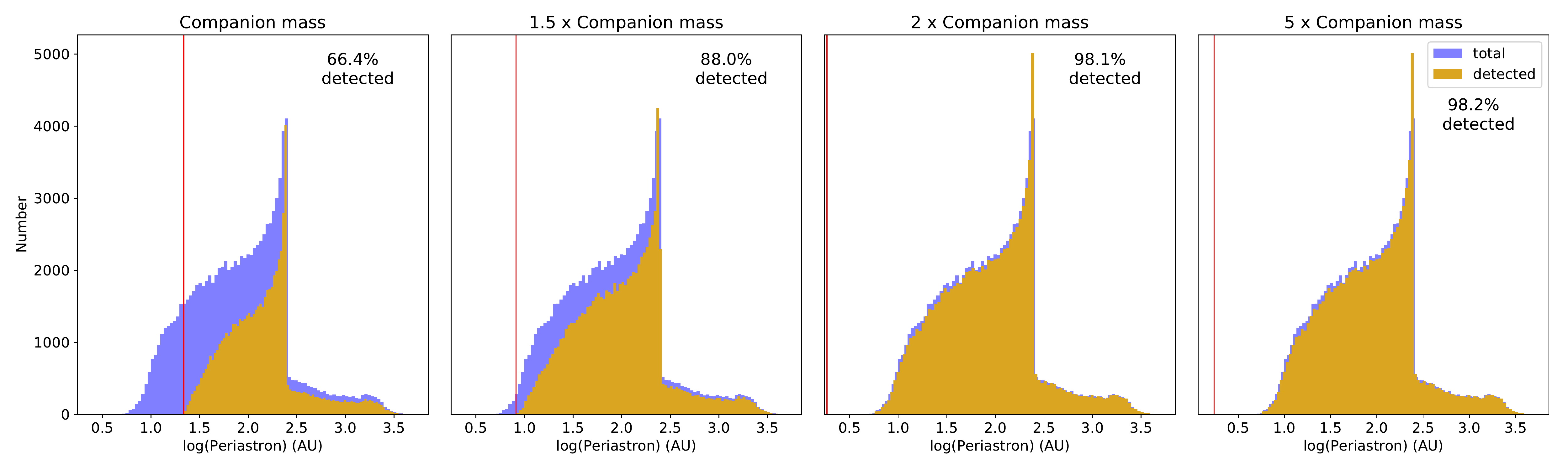}
\caption{\small{Top: Number of scatterers that would have been detectable for a scatterer of equal mass, 1.5 times companion's mass, twice the companion's mass, and five times the companion's mass, plotted as a function of companion's periastron distance, assuming both objects were on an initially circular orbit at the companion's current periastron.  Each plot shows the number of detected scatterers (gold) over-plotted atop the number of total orbits (purple).  The detection limit for each mass is plotted as a red vertical line.  In the vast majority of cases, a second companion of equal or greater mass would have been detected in existing observations, so it is unlikely for GSC 6214-210\,b to have been scattered from close radii, unless it both originated from deep in the potential well and was scattered by a similar-mass companion.  Bottom: Fraction of detected scatterers in the most limiting energy case, a scatterer orbit with eccentricity\;=\;1 and apastron at the companion's current periastron distance.  We see that in this most conservative case, the fraction of detected scatterers is smaller, but the overall conclusion of the simulation remains the same --- in the majority of cases the scatterer would have been detected in imaging.  }}
\label{fig:detected_scatters}
\end{figure*}

Figure \ref{fig:detected_scatters} shows histograms of the number of detected and undetected scatterers as a function of the companion's current periastron distance for both scattering exercises.  In the most conservative case, for scatterers of mass equal to the companion, 66.4\% of potential scatterers would have been detected in imaging.  However, that 66.4\% comprises the ones on wider orbits; orbits fully or mostly inside the detection limit have little chance of being observed.  So we cannot rule out the existence of a second companion, except in the highest mass cases.  But, in the majority of cases, our conservative simulation suggests that a scattering object would have been seen in the existing observations.  We therefore conclude that GSC 6214-210 b was unlikely to have been scattered from an initially tighter orbit, unless it was scattered from a very tight orbit and only by a scatterer of similar mass. 

%%%%%%%%%%%%%%%%%%%%%%%%%%%%%%%%% discussion %%%%%%%%%%%%%%%%%%%%%%%%%%%%%%%%%%%%%

\section{Discussion} \label{sec:discussion}
We have used the orbit of the companion to look for clues regarding its formation mechanism, specifically determining the likelihood of formation at a close radius followed by ejection via scattering.  Here we present evidence that dynamical scattering is a possible but not likely explanation for the formation of this object.

Unlike some other studies of wide PMC orbits (e.g., \citealt{Bryan}), this study was unable to entirely rule out high eccentricity orbits with close periastron.  Orbits with eccentricity near 0.9 are common in our posterior, and orbits with periastron less than 10 AU have high likelihoods.  We cannot rule out dynamical scattering based on orbital parameter posteriors alone. 

While orbits with periastron less than 10 AU comprise about 2\% of the posterior, orbits with periastron less than 5 AU (0.03\% of all orbits), all have low statistical likelihoods.  So, we conclude that orbits which would have allowed the companion to form inside of Jupiter's orbit, and near ice lines at $\sim$1-2 AU, are not preferred. 
However, orbits with periastron 5 - 30 AU are common in our sample (21\%), many with high statistical likelihoods, so formation of this companion within that range cannot be ruled out by this analysis alone.  

Other lines of evidence support the conclusion that dynamical scattering is unlikely to explain the origin of this companion.  \citet{Bowler2014} observed GSC 6214-210\,b to be actively accreting from a circumplanetary disk.  They determined that disruption of circumplanetary disks in scattering interactions is likely quite common, and interpret the presence of a disk as evidence against a past scattering event.

Dynamical scattering from a close radius would require another companion of equal or larger mass to act as the scatterer.
\citet{Bryan} conducted a direct-imagining campaign to search for close-in scatterers on seven known wide PMC systems, and were unable to detect any new companions, and concluded that dynamical scattering was not a good explanation for the formation of those systems.  We performed a simulation of the ability to detect potential scatterers in imaging data for GSC 6214-210, and found that for all scattering objects of equal and greater mass, the scatterer would have been observed in the majority of cases.  Simulations of scatterers of equal mass show it would not have been detected in a relatively high fraction of cases (32\% for equal mass, 23.5\% for 1.5 x companion mass), so we cannot conclusively rule out the existence of a second close-in companion.  However, our simulations represented the most conservative case, so we conclude that a second scatterer would have been observed in the majority of images.  No other objects have been detected in ten years of imaging data, so we interpret this as evidence against the existence of such an object.  Additionally, \textit{Gaia} DR2 reported an astrometric excess noise of 0.0, implying absence of an invisible companion large enough to scatter GSC\,6214-210\,b and supporting the conclusion that an inner companion is unlikely.
 
While we cannot conclusively rule out dynamical scattering, these lines of evidence form a picture for which scattering is not a likely explanation for this PMC.  This conclusion is in agreement with the dominant body of work for wide-orbit planet formation,  discussed in Section \ref{sec:intro}, and continues to point to in-situ formation at the more likely scenario.
Measuring the orbital architecture of additional wide PMC systems would provide more evidence to support or challenge this conclusion.

%%%%%%%%%%%%%%%%%%%%%%%%%%%%%%%%% conclusion %%%%%%%%%%%%%%%%%%%%%%%%%%%%%%%%%%%%%

\section{Summary}

We conducted an astrometric study and orbital motion determination for the wide-orbit planetary mass companion GSC 6214-210 b, yielding the first statistically significant measurement of orbital motion for this object.  We used a Gaussian PSF model with an MCMC to measure relative astrometry in ten years of AO imaging data from NIRC2 at Keck, and then developed a custom implementation of the Orbits for the Impatient rejection sampling algorithm \citep{blunt} to fit orbital parameters. 

Our results demonstrate that a wide range of orbital parameters fit the data, including some with low periastron distances. However, we also demonstrate that the tightest orbits still produce a relatively poor fit, with low likelihood to have been observed. We also demonstrate that for most of the allowed orbits, the scattering object would likely have been detected in past high-resolution imaging data, unless the companion was scattered from the tightest orbits and specifically by a scatterer of similar mass.
We cannot rule out core accretion with subsequent dynamical scattering as a viable formation pathway for this companion, however we conclude that it is much less likely than in-situ mechanisms such as gravitational instability of the protostellar disk.  This is in agreement with prior work on other wide PMC systems, and suggests that scattering is not likely to play a key role in producing PMCs.   

\acknowledgments

The data presented herein were obtained at the W. M. Keck Observatory, which is operated as a scientific partnership among the California Institute of Technology, the University of California and the National Aeronautics and Space Administration. The Observatory was made possible by the generous financial support of the W. M. Keck Foundation.  This research has also made use of the Keck Observatory Archive (KOA), which is operated by the W. M. Keck Observatory and the NASA Exoplanet Science Institute (NExScI), under contract with the National Aeronautics and Space Administration.  The authors wish to recognize and acknowledge the very significant cultural role and reverence that the summit of Maunakea has always had within the indigenous Hawaiian community.  We are most fortunate to have the opportunity to conduct observations from this mountain.

L.A.P. was supported by a NASA/Keck Data Analysis Grant and by the McDonald Observatory Board of Visitors, the Cox Endowment Fund of the UT-Austin Department of Astronomy, the Barry Goldwater Scholarship and Excellence in Education Foundation, and the Astronaut Scholarship Foundation.

T.J.D. acknowledges research support from Gemini Observatory.

M.J.I. was supported by the Australian Research Council Future Fellowship (FT130100235)

E.K.B. gratefully acknowledges the support of the Australian Government Research Training Program Stipend Scholarship. 

L.A.P. acknowledges the contributions of Richard Seifert, without whose consistent and enthusiastic coding support this project would have suffered. She also wishes to thank Adam Kraus, Trent Dupuy, and Brendan Bowler for their mentorship and generosity in giving their time towards this project.

The authors acknowledge the Texas Advanced Computing Center (TACC) at The University of Texas at Austin for providing high performance computing resources that have contributed to the research results reported within this paper. URL: http://www.tacc.utexas.edu.

This work has made use of data from the European Space Agency (ESA)
mission {\it Gaia} (\url{https://www.cosmos.esa.int/gaia}), processed by
the {\it Gaia} Data Processing and Analysis Consortium (DPAC,
\url{https://www.cosmos.esa.int/web/gaia/dpac/consortium}). Funding
for the DPAC has been provided by national institutions, in particular
the institutions participating in the {\it Gaia} Multilateral Agreement.

This research has made use of the SIMBAD database,
operated at CDS, Strasbourg, France

\facility{Keck:II (NIRC2), Texas Advanced Computing Center (TACC)} 

\software{StarFinder \citep{diolaiti2000}, Numpy \citep{numpy}, Astropy \citep{astropy:2018}, Matplotlib \citep{Hunter:2007}}

%%%%%%%%%%%%%%%%%%%%%%%%%%%%%%%%%%%%%%%%%%%%%%%%%%%%%%%%%%%%%%%%%%%%%%%%%%%%

\appendix
%\section{Appendix}
\section{Detection Limits}\label{detection limits}

Over the 8 years since we reported the results of Ireland et al. (2011) we have observed the GSC 6214-210 system numerous additional times with both imaging and nonredundant mask (NRM) interferometry, while also improving our understanding of the noise properties of the NIRC2 camera. We therefore are now able to provide updated detection limits for additional companions interior to GSC 6214-210\,b.

%For each of the 8 imaging 
For the 2016 epoch that we describe in Section \ref{sec:obs}, we conducted PSF subtraction and candidate identification using the methods described by Kraus et al. (2016). To briefly summarize, each image 
%of each epoch 
was first calibrated and cleaned of hot pixels and cosmic rays. To determine optimal detection limits at wide separations, we then subtracted the azimuthally-averaged flux profile of the primary star, computed the flux at every integer pixel location through a photometric aperture of diameter $\lambda /D$, and searched for any such flux values that were $+6\sigma$ outliers compared to the distribution of all such values at that projected separation. To search for companions at smaller separations, we then repeated the same procedure while subtracting the best-fitting PSF as selected from all observations of apparently single stars within the NIRC2 archive, based on the $\chi^2$ fit at radii of 0.15--0.45\arcsec. 
%In all cases, 
We did not find any candidate detections at $>+6\sigma$, establishing an upper limit for additional companions at the corresponding flux.

For the 3 NRM epochs that we have taken of GSC 6214-210 (one in the L$^{\prime}$ filter, two in the CH4S filter), we analyzed the data using the pipeline described by Kraus et al. (2008, 2011). To briefly summarize, the data analysis takes three broad steps: basic image analysis (flatfielding, bad pixel removal, dark subtraction), extraction and calibration of squared visibilities and closure phases, and binary model fitting. Unless testing fits for close, near-equal binaries, we fit only to closure phase, as this is the quantity most robust to changes in the AO point-spread function. The detection limits are found using a Monte-Carlo method that simulates 10,000 random closure-phase datasets of a point source, with closure-phase errors and covariances that match the calibrated target data set. This routine then searches for the best fit for a companion in each randomized dataset. Over each annulus of projected separation from the priamry star, the 99.9\% confidence limit is set to the contrast ratio where 99.9\% of the Monte-Carlo trials have no best binary fit with a companion brighter than this limit anywhere within the annulus. We did not find any cases where a companion was detected above that limit in a dataset, so we adopt those contrast values as our limits on the existence of additional companions in the system.

In Table \ref{table:imaging limits}, we list the detection limits for additional companions in the 2016 dataset, and in Table \ref{table:nrm limits}, we similarly list the detection limits for additional companions in each NRM dataset. The deepest example of each dataset was used as an input for the limits on potential scatterers, as we describe in Section \ref{sec:scatterers}.

\begin{deluxetable*}{cccccccccccccc}[htb!]
\tablecaption{{Contrast limits for NIRC2 imaging of GSC 6214-210}\label{table:imaging limits}}
%\tablewidth{0.45\textwidth}
\tablehead{\colhead{Epoch} & \colhead{MJD} & \colhead{Number of} & \colhead{Integration} & \multicolumn{10}{c}{Contrast Limit ($\Delta K^{\prime}$ in mag) at Projected Separation ($\rho$ in mas)}\\
 & & \colhead{Frames} & \colhead{Time (s)} & \colhead{150} & \colhead{200}& \colhead{250}& \colhead{300} & \colhead{400}& \colhead{500}& \colhead{700}& \colhead{1000} & \colhead{1500} & \colhead{2000}}
\startdata
2016.46 & 57555.47 & 10 & 20 & 5.0 & 5.9 & 6.3 & 6.6 & 7.1 & 7.3 & 8.1 & 8.3 & 8.3 & 8.2 \\
\enddata
%\tablecomments{ }
\end{deluxetable*}

\begin{deluxetable*}{cccccccc}[htb!]
\tablecaption{{NRM Contrast limits for GSC 6214-210}\label{table:nrm limits}}
%\tablewidth{0.45\textwidth}
\tablehead{\colhead{Epoch} & \colhead{MJD} & \colhead{Filter} &  \colhead{Number of}  & \multicolumn{3}{c}{Limits (mag) at}\\
 & & & \colhead{Frames}  & \colhead{10 mas} & \colhead{20 mas}& \colhead{40 mas}}
\startdata
2011.42 & 55716.34 & L$^{\prime}$ & 6 & 0.27 &  3.20 &  3.27\\
2014.58 & 56868.27 & CH4S & 6 & 1.89 & 4.42 & 4.46\\
2017.49 & 57932.31 & CH4S & 13 & 0.27 &  3.20 &  3.27\\
\enddata
%\tablecomments{ }
\end{deluxetable*}

\clearpage

\section{Orbital Parameters Joint Credible Intervals}\label{corner plot}
\begin{figure*}[htb!]
\centering
\includegraphics[width=0.90\textwidth]{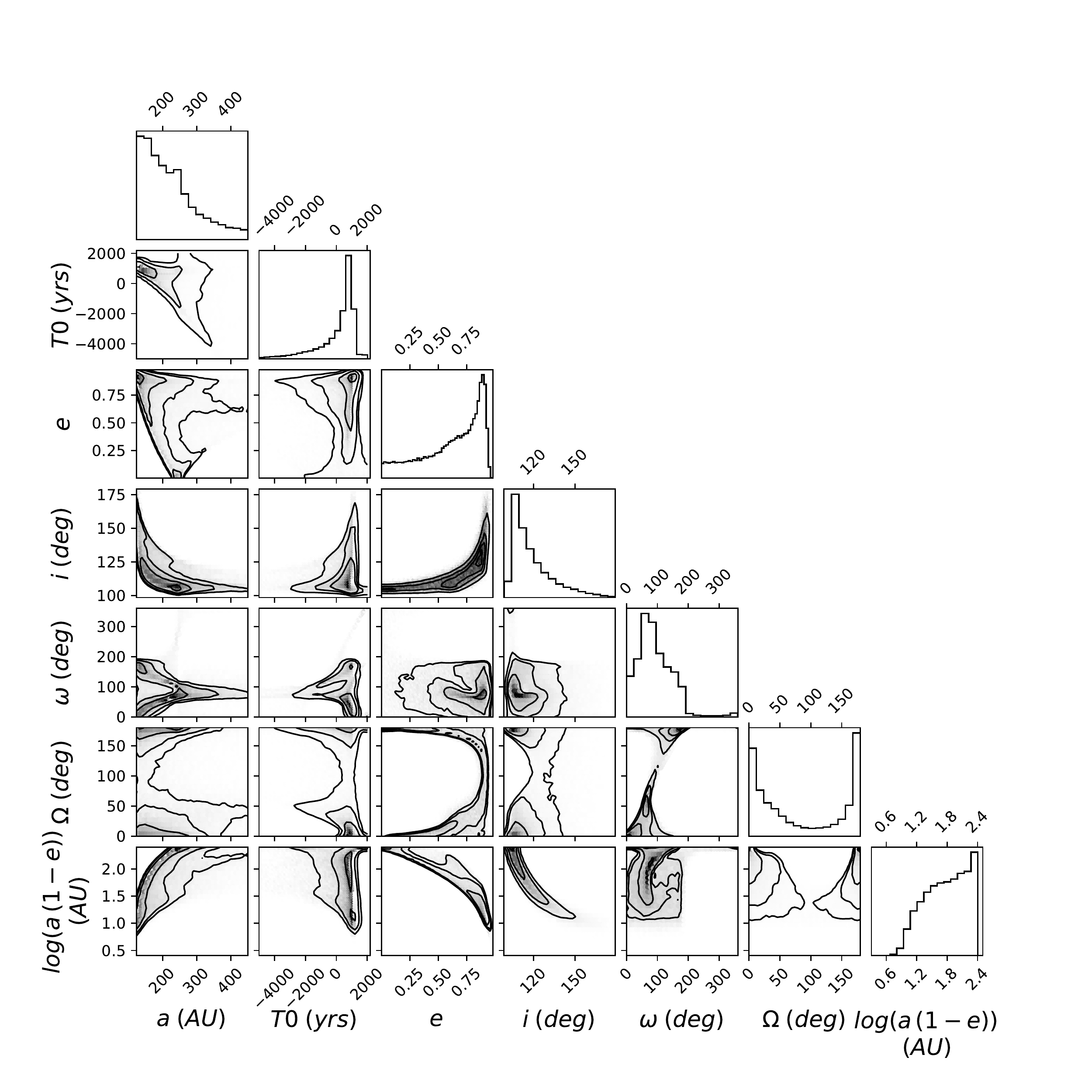}
\caption{\small{Two-dimensional marginalized credible intervals for the orbital elements for the companion to GSC 6214-210.  The one-dimensional marginalized distributions of each element is shown on the diagonal.  Parameters plotted are semi-major axis (a) in AU, epoch of periastron passage (To), eccentricity (e), inclination (i) in degrees, argument of periastron ($\omega$) in degrees, and longitude of periastron ($\Omega$) in degrees, and periastron distance ((1-e) a) in AU.  The contours in the histograms correspond to the 1-, 2-, 3-, and 4-$\sigma$ levels.}}
\label{fig:corner}
\end{figure*}

Figure \ref{fig:corner} reports the full joint credible intervals for the fit parameters, plus periastron, for the unconstrained orbit fit.

\clearpage

\section{Image Astrometry}\label{images appendix}
In Table \ref{table:astrometry}, we report the observational details and our individual fit results for each of the images we consider in this work. The individual measurements were combined into a single measurement at each epoch (Table \ref{table:summary}) using the methods described in Section \ref{sec:relative astrometry}.

\startlongtable
\begin{deluxetable*}{cccccccc}
\tablecaption{{Astrometric Measurements by Image}\label{table:astrometry}}
\tablehead{
\colhead{Image} & \colhead{t\s{int}} & \colhead{Tracking} & \colhead{$\rho$} & \colhead{PA} & \colhead{S/N} & \colhead{S/N} & \colhead{FWHM} \\
\colhead{Filename} & \colhead{(sec)} & \colhead{Mode} & \colhead{(mas)} & \colhead{(deg)} & \colhead{Primary} & \colhead{Companion} & \colhead{(mas)}
}
\startdata
\rowgroup{2008 Jun 17 (JD 2454634.5) & PI: Ireland} \\
N2.20080617.29357.fits & 20.0 & PA & 2202.23 $\pm \,0.57$ & 175.624 $\pm \,0.015$ & 1836 & 105 & 44.51 $\pm \,0.02$ \\
N2.20080617.29399.fits & 20.0 & PA & 2202.66 $\pm \,0.58$ & 175.607 $\pm \,0.016$ & 2103 & 120 & 44.81 $\pm \,0.02$ \\
N2.20080617.29442.fits & 20.0 & PA & 2201.82 $\pm \,0.64$ & 175.624 $\pm \,0.017$ & 2075 & 100 & 44.44 $\pm \,0.02$ \\
N2.20080617.29484.fits & 20.0 & PA & 2202.01 $\pm \,0.57$ & 175.619 $\pm \,0.015$ & 2155 & 109 & 44.74 $\pm \,0.02$ \\
N2.20080617.29527.fits & 20.0 & PA & 2202.87 $\pm \,0.62$ & 175.616 $\pm \,0.017$ & 2068 & 100 & 46.13 $\pm \,0.02$ \\
N2.20080617.29570.fits & 20.0 & PA & 2201.65 $\pm \,0.67$ & 175.615 $\pm \,0.018$ & 2041 & 109 & 46.01 $\pm \,0.02$ \\
N2.20080617.29615.fits & 20.0 & PA & 2201.68 $\pm \,0.57$ & 175.607 $\pm \,0.015$ & 1945 & 109 & 44.56 $\pm \,0.02$ \\
N2.20080617.29658.fits & 20.0 & PA & 2201.20 $\pm \,0.60$ & 175.616 $\pm \,0.016$ & 2089 & 113 & 44.81 $\pm \,0.02$ \\
N2.20080617.29701.fits & 20.0 & PA & 2203.52 $\pm \,0.62$ & 175.618 $\pm \,0.017$ & 2198 & 102 & 45.09 $\pm \,0.02$ \\
N2.20080617.29744.fits & 20.0 & PA & 2201.69 $\pm \,0.55$ & 175.624 $\pm \,0.014$ & 1591 & 120 & 41.03 $\pm \,0.01$ \\
N2.20080617.29818.fits & 20.0 & PA & 2202.17 $\pm \,0.54$ & 175.639 $\pm \,0.014$ & 1555 & 117 & 42.40 $\pm \,0.02$ \\
N2.20080617.29860.fits & 20.0 & PA & 2201.35 $\pm \,0.50$ & 175.630 $\pm \,0.013$ & 1516 & 137 & 40.86 $\pm \,0.02$ \\
N2.20080617.29903.fits & 20.0 & PA & 2200.83 $\pm \,0.52$ & 175.632 $\pm \,0.013$ & 1646 & 124 & 41.10 $\pm \,0.02$ \\
N2.20080617.29945.fits & 20.0 & PA & 2202.16 $\pm \,0.53$ & 175.629 $\pm \,0.014$ & 1496 & 136 & 40.65 $\pm \,0.02$ \\
N2.20080617.29988.fits & 20.0 & PA & 2201.70 $\pm \,0.55$ & 175.631 $\pm \,0.014$ & 1561 & 136 & 43.81 $\pm \,0.02$ \\
N2.20080617.30031.fits & 20.0 & PA & 2202.18 $\pm \,0.56$ & 175.615 $\pm \,0.014$ & 1359 & 118 & 42.01 $\pm \,0.02$ \\
N2.20080617.30074.fits & 20.0 & PA & 2202.19 $\pm \,0.56$ & 175.633 $\pm \,0.014$ & 1368 & 115 & 41.50 $\pm \,0.02$ \\
N2.20080617.30117.fits & 20.0 & PA & 2201.80 $\pm \,0.55$ & 175.620 $\pm \,0.014$ & 1310 & 126 & 41.07 $\pm \,0.02$ \\
N2.20080617.30160.fits & 20.0 & PA & 2201.93 $\pm \,0.51$ & 175.626 $\pm \,0.013$ & 1162 & 138 & 41.22 $\pm \,0.01$ \\
N2.20080617.30203.fits & 20.0 & PA & 2201.18 $\pm \,0.59$ & 175.629 $\pm \,0.015$ & 1277 & 126 & 41.72 $\pm \,0.02$ \\
Mean: & & & 2201.96 $\pm \,1.09$ & 175.623 $\pm \,0.027$ \\
\hline
\rowgroup{2009 May 31 (JD 2454982.88) & PI: Hillenbrand} \\
N2.20090531.32959.fits & 10.0 & PA & 2202.32 $\pm \,1.51$ & 175.566 $\pm \,0.038$ & 1890 & 46 & 40.77 $\pm \,0.03$ \\
N2.20090531.33040.fits & 10.0 & PA & 2203.08 $\pm \,1.43$ & 175.575 $\pm \,0.033$ & 2459 & 10 & 39.25 $\pm \,0.03$ \\
N2.20090531.33067.fits & 10.0 & PA & 2203.84 $\pm \,1.39$ & 175.556 $\pm \,0.033$ & 2455 & 10 & 39.20 $\pm \,0.03$ \\
N2.20090531.33115.fits & 10.0 & PA & 2199.76 $\pm \,1.73$ & 175.597 $\pm \,0.040$ & 2276 & 46 & 40.23 $\pm \,0.03$ \\
N2.20090531.33141.fits & 10.0 & PA & 2199.90 $\pm \,1.69$ & 175.610 $\pm \,0.040$ & 2240 & 49 & 40.41 $\pm \,0.03$ \\
N2.20090531.33190.fits & 10.0 & PA & 2204.76 $\pm \,1.44$ & 175.605 $\pm \,0.034$ & 2109 & 48 & 38.72 $\pm \,0.03$ \\
N2.20090531.33216.fits & 10.0 & PA & 2205.79 $\pm \,1.39$ & 175.595 $\pm \,0.032$ & 2240 & 46 & 38.61 $\pm \,0.03$ \\
N2.20090531.33310.fits & 10.0 & PA & 2200.63 $\pm \,1.50$ & 175.584 $\pm \,0.039$ & 2043 & 54 & 40.77 $\pm \,0.03$ \\
N2.20090531.33336.fits & 10.0 & PA & 2200.69 $\pm \,1.55$ & 175.583 $\pm \,0.040$ & 1885 & 44 & 40.77 $\pm \,0.03$ \\
Mean: & & & 2202.56 $\pm \,1.20$ & 175.585 $\pm \,0.031$ \\
\hline
\rowgroup{2010 Apr 26 (JD 2455313.11) & PI: Kraus} \\
N2.20100426.52760.fits & 10.0 & PA & 2201.54 $\pm \,0.58$ & 175.545 $\pm \,0.013$ & 3238 & 51 & 37.99 $\pm \,0.02$ \\
N2.20100426.52820.fits & 10.0 & PA & 2200.67 $\pm \,0.54$ & 175.590 $\pm \,0.012$ & 2781 & 136 & 38.53 $\pm \,0.02$ \\
N2.20100426.52838.fits & 10.0 & PA & 2201.52 $\pm \,0.60$ & 175.550 $\pm \,0.013$ & 2867 & 129 & 38.69 $\pm \,0.02$ \\
N2.20100426.52885.fits & 10.0 & PA & 2201.75 $\pm \,0.55$ & 175.570 $\pm \,0.012$ & 2710 & 145 & 37.86 $\pm \,0.02$ \\
N2.20100426.52902.fits & 10.0 & PA & 2201.99 $\pm \,0.62$ & 175.636 $\pm \,0.013$ & 2782 & 115 & 38.11 $\pm \,0.01$ \\
N2.20100426.52947.fits & 10.0 & PA & 2200.75 $\pm \,1.03$ & 175.584 $\pm \,0.025$ & 1503 & 101 & 65.30 $\pm \,0.03$ \\
N2.20100426.52965.fits & 10.0 & PA & 2202.33 $\pm \,0.90$ & 175.602 $\pm \,0.021$ & 1698 & 115 & 58.86 $\pm \,0.06$ \\
N2.20100426.53014.fits & 10.0 & PA & 2200.63 $\pm \,0.71$ & 175.562 $\pm \,0.015$ & 1681 & 110 & 61.09 $\pm \,0.04$ \\
Mean: & & & 2201.40 $\pm \,1.40$ & 175.578 $\pm \,0.036$ \\
\hline
\rowgroup{2011 Jun 04 (JD 2455716.84) & PI: Kraus} \\
N2.20110604.29500.fits & 10.0 & PA & 2200.26 $\pm \,1.60$ & 175.488 $\pm \,0.032$ & 1887 & 61 & 41.16 $\pm \,0.03$ \\
N2.20110604.29538.fits & 10.0 & PA & 2200.18 $\pm \,1.37$ & 175.489 $\pm \,0.028$ & 1835 & 63 & 40.63 $\pm \,0.03$ \\
N2.20110604.29597.fits & 10.0 & PA & 2204.18 $\pm \,1.70$ & 175.534 $\pm \,0.031$ & 1283 & 67 & 40.60 $\pm \,0.03$ \\
N2.20110604.29625.fits & 10.0 & PA & 2204.58 $\pm \,1.76$ & 175.521 $\pm \,0.033$ & 1310 & 54 & 41.45 $\pm \,0.04$ \\
N2.20110604.29660.fits & 10.0 & PA & 2201.78 $\pm \,1.60$ & 175.521 $\pm \,0.031$ & 1442 & 61 & 39.68 $\pm \,0.03$ \\
N2.20110604.29724.fits & 10.0 & PA & 2204.29 $\pm \,1.22$ & 175.493 $\pm \,0.024$ & 1754 & 89 & 39.55 $\pm \,0.03$ \\
Mean: & & & 2202.53 $\pm \,1.12$ & 175.506 $\pm \,0.026$ \\
\hline
\rowgroup{2014 Jul 30 (JD 2456868.77) & PI: Dupuy} \\
N2.20140730.23631.fits & 20.0 & VA & 2195.00 $\pm \,3.22$ & 175.449 $\pm \,0.091$ & 1743 & 22 & 38.69 $\pm \,0.06$ \\
N2.20140730.23677.fits & 20.0 & VA & 2200.10 $\pm \,3.47$ & 175.479 $\pm \,0.084$ & 1351 & 30 & 51.15 $\pm \,0.11$ \\
N2.20140730.23723.fits & 20.0 & VA & 2199.07 $\pm \,1.34$ & 175.455 $\pm \,0.036$ & 2136 & 55 & 41.23 $\pm \,0.03$ \\
N2.20140730.23785.fits & 20.0 & VA & 2197.70 $\pm \,0.87$ & 175.461 $\pm \,0.025$ & 3220 & 92 & 40.04 $\pm \,0.02$ \\
N2.20140730.23831.fits & 20.0 & VA & 2197.34 $\pm \,0.83$ & 175.457 $\pm \,0.022$ & 3719 & 87 & 44.72 $\pm \,0.02$ \\
N2.20140730.23877.fits & 20.0 & VA & 2197.28 $\pm \,0.90$ & 175.436 $\pm \,0.023$ & 3307 & 77 & 38.58 $\pm \,0.02$ \\
Mean: & & & 2197.63 $\pm \,0.64$ & 175.452 $\pm \,0.015$ \\
\hline 
\rowgroup{2016 Jun 16 (JD 2457555.97) & PI: Ireland} \\
N2.20160616.40557.fits & 20.0 & VA & 2200.34 $\pm \,4.52$ & 175.493 $\pm \,0.105$ & 1559 & 18 & 40.20 $\pm \,0.05$ \\
N2.20160616.40643.fits & 20.0 & VA & 2197.54 $\pm \,4.81$ & 175.405 $\pm \,0.134$ & 1380 & 16 & 47.76 $\pm \,0.10$ \\
N2.20160616.40712.fits & 20.0 & VA & 2197.90 $\pm \,2.56$ & 175.411 $\pm \,0.073$ & 1495 & 21 & 51.45 $\pm \,0.09$ \\
N2.20160616.40755.fits & 20.0 & VA & 2198.65 $\pm \,2.78$ & 175.459 $\pm \,0.087$ & 1294 & 23 & 50.22 $\pm \,0.08$ \\
N2.20160616.40798.fits & 20.0 & VA & 2196.86 $\pm \,1.72$ & 175.438 $\pm \,0.048$ & 1647 & 37 & 47.42 $\pm \,0.06$ \\
N2.20160616.40841.fits & 20.0 & VA & 2197.91 $\pm \,1.39$ & 175.421 $\pm \,0.041$ & 1635 & 35 & 47.97 $\pm \,0.06$ \\
N2.20160616.40910.fits & 20.0 & VA & 2198.96 $\pm \,1.66$ & 175.451 $\pm \,0.047$ & 1616 & 35 & 52.75 $\pm \,0.07$ \\
N2.20160616.40953.fits & 20.0 & VA & 2199.69 $\pm \,1.49$ & 175.454 $\pm \,0.044$ & 1652 & 44 & 50.60 $\pm \,0.06$ \\
N2.20160616.40996.fits & 20.0 & VA & 2197.91 $\pm \,1.64$ & 175.368 $\pm \,0.052$ & 1509 & 35 & 55.11 $\pm \,0.10$ \\
N2.20160616.41039.fits & 20.0 & VA & 2197.83 $\pm \,1.35$ & 175.443 $\pm \,0.043$ & 1517 & 38 & 51.91 $\pm \,0.07$ \\
Mean: & & & 2198.24 $\pm \,1.10$ & 175.433 $\pm \,0.026$ \\
\hline
\rowgroup{2017 Jun 28 (JD 2457932.81) & PI: Kraus} \\
N2.20170628.26610.fits & 20.0 & PA & 2198.29 $\pm \,0.94$ & 175.364 $\pm \,0.023$ & 708 & 74 & 41.54 $\pm \,0.02$ \\
N2.20170628.26653.fits & 20.0 & PA & 2198.48 $\pm \,1.08$ & 175.386 $\pm \,0.026$ & 835 & 67 & 41.38 $\pm \,0.03$ \\
N2.20170628.26728.fits & 20.0 & PA & 2197.70 $\pm \,1.10$ & 175.391 $\pm \,0.028$ & 764 & 65 & 45.83 $\pm \,0.02$ \\
N2.20170628.26771.fits & 20.0 & PA & 2196.93 $\pm \,0.78$ & 175.394 $\pm \,0.020$ & 701 & 75 & 44.50 $\pm \,0.02$ \\
N2.20170628.26872.fits & 20.0 & PA & 2197.26 $\pm \,0.57$ & 175.357 $\pm \,0.014$ & 609 & 92 & 42.12 $\pm \,0.02$ \\
N2.20170628.26915.fits & 20.0 & PA & 2196.53 $\pm \,0.60$ & 175.354 $\pm \,0.015$ & 643 & 106 & 42.29 $\pm \,0.01$ \\
N2.20170628.26958.fits & 20.0 & PA & 2198.91 $\pm \,1.02$ & 175.375 $\pm \,0.027$ & 708 & 50 & 45.87 $\pm \,0.03$ \\
N2.20170628.27001.fits & 20.0 & PA & 2198.30 $\pm \,1.15$ & 175.413 $\pm \,0.033$ & 565 & 37 & 49.66 $\pm \,0.04$ \\
N2.20170628.27086.fits & 20.0 & PA & 2198.02 $\pm \,0.82$ & 175.333 $\pm \,0.021$ & 556 & 65 & 43.60 $\pm \,0.02$ \\
N2.20170628.27186.fits & 20.0 & PA & 2196.83 $\pm \,0.83$ & 175.344 $\pm \,0.021$ & 514 & 76 & 44.26 $\pm \,0.02$ \\
N2.20170628.27229.fits & 20.0 & PA & 2196.48 $\pm \,0.66$ & 175.387 $\pm \,0.017$ & 545 & 99 & 43.86 $\pm \,0.02$ \\
N2.20170628.27272.fits & 20.0 & PA & 2195.19 $\pm \,0.60$ & 175.351 $\pm \,0.016$ & 499 & 83 & 43.15 $\pm \,0.02$ \\
N2.20170628.27315.fits & 20.0 & PA & 2197.82 $\pm \,0.68$ & 175.399 $\pm \,0.017$ & 503 & 89 & 42.77 $\pm \,0.02$ \\
Mean: & & & 2197.11 $\pm \,1.13$ & 175.368 $\pm \,0.027$ \\
\hline
\rowgroup{2018 Jul 01 (JD 2458300.87) & PI: Ireland} \\
N2.20180701.31632.fits & 20.0 & PA & 2194.63 $\pm \,0.62$ & 175.363 $\pm \,0.014$ & 2534 & 104 & 38.62 $\pm \,0.01$ \\
N2.20180701.31675.fits & 20.0 & PA & 2195.94 $\pm \,0.67$ & 175.375 $\pm \,0.015$ & 1897 & 91 & 37.64 $\pm \,0.01$ \\
N2.20180701.31718.fits & 20.0 & PA & 2196.54 $\pm \,0.61$ & 175.384 $\pm \,0.014$ & 1694 & 88 & 38.18 $\pm \,0.01$ \\
N2.20180701.31761.fits & 20.0 & PA & 2196.01 $\pm \,0.63$ & 175.389 $\pm \,0.015$ & 1822 & 91 & 39.05 $\pm \,0.01$ \\
N2.20180701.31804.fits & 20.0 & PA & 2195.15 $\pm \,0.60$ & 175.365 $\pm \,0.014$ & 2199 & 105 & 38.63 $\pm \,0.01$ \\
Mean: & & & 2195.64 $\pm \,1.12$ & 175.375 $\pm \,0.027$ \\
\enddata

\tablecomments{All images were obtained with the NIRC2 imager on Keck II in the K$^{\prime}$ filter with no coronagraph.  Integration time (t\s{int}) is determined as integration time per coadd times the number of coadds. Tracking mode is either Position Angle Mode (PA) or Vertical Angle Mode (VA).  SNR is computed as a ratio of the sky-subtracted sum of pixel values within an aperture centered over the object to the standard deviation of pixel values in an annulus outside the object's PSF. FWHM is determined as 2.355 x the larger of the two sigmas of the 2D Gaussian fit.}

\end{deluxetable*}

\bibliographystyle{yahapj}
\bibliography{references}

\(\)

\end{document}